\documentclass[10pt,journal,compsoc]{IEEEtran}
\usepackage{amsmath}
\usepackage{graphicx}
\usepackage[colorinlistoftodos]{todonotes}
\usepackage{amssymb,amsmath,amsthm,epsfig,url,mathrsfs} 
\usepackage[framemethod=TikZ]{mdframed}
\usepackage{xcolor}[dvipsnames]
\usepackage{ragged2e,comment}
\usepackage{amsmath,mathrsfs}
\usepackage{subcaption}
\usepackage{booktabs}
\usepackage{caption}
\usepackage{subcaption}
\usepackage{colortbl}
\usepackage{hyperref}

\renewcommand{\vec}[1]{\boldsymbol{\mathrm{#1}}}

\usepackage{multirow}
\usepackage{url,lipsum,cite}

\theoremstyle{definition}

\newtheorem{theorem}{Theorem}[section]
\newtheorem{lemma}[theorem]{Lemma}

\newcommand{\mydef}{\ensuremath{\triangleq}}
\renewcommand{\vec}{\boldsymbol} 

\newcommand\DBprior{\pi}
\definecolor{DarkGreen}{RGB}{1,50,32}

\definecolor{DarkGreen}{rgb}{0.0, 0.2, 0.13}

\usepackage{pgfplots}
\pgfplotsset{compat=1.14}
\usetikzlibrary{decorations.pathreplacing,angles,quotes}
\usetikzlibrary{patterns,patterns.meta}
\usepgfplotslibrary{colorbrewer}

\hypersetup{
   unicode=false,          
   pdftoolbar=true,        
   pdfmenubar=true,        
   pdffitwindow=false,     
   pdfstartview={FitH},    
   pdftitle={My title},    
   pdfauthor={Author},     
   pdfsubject={Subject},   
   pdfcreator={Creator},   
   pdfproducer={Producer}, 
   pdfkeywords={keyword1} {key2} {key3}, 
   pdfnewwindow=true,      
   colorlinks=true,       
   linkcolor=blue,          
   citecolor=blue,        
   filecolor=magenta,      
   urlcolor=blue           
}
\usepackage{hyperref}

\ifCLASSINFOpdf
\else
\fi
\begin{document}

\thispagestyle{empty}
\onecolumn

\begin{center}
    {\huge\textbf{t-EER: Parameter-Free Tandem Evaluation of\\\vspace{0.5ex} Countermeasures and Biometric Comparators}}\\
    \vspace{3ex}
    Tomi H. Kinnunen, Kong Aik Lee, Hemlata Tak, Nicholas Evans and Andreas Nautsch
\end{center}

\vspace{2ex}
\noindent This paper has been accepted for publication in \emph{IEEE Transactions on Pattern Analysis and Machine Intelligence}. If you would like to cite the work, please use the following bibliographic entry.
\vspace{5ex}

\texttt{   T. Kinnunen, K.A. Lee, H. Tak, N. Evans and A. Nautsch, "t-EER: Parameter-Free Tandem 
                    Evaluation of Countermeasures and Biometric Comparators", to appear in IEEE Transactions on Pattern Analysis and Machine Intelligence (2023), doi: 10.1109/TPAMI.2023.3313648}
\vspace{3ex}

\begin{verbatim}
    @ARTICLE {Kinnunen2023-tEER,
        author    = {T. Kinnunen and K.A. Lee and H. Tak and N. Evans and A. Nautsch},
        journal   = {{IEEE} Transactions on Pattern Analysis and Machine Intelligence},
        title     = {t-EER: Parameter-Free Tandem Evaluation of Countermeasures and 
                    Biometric Comparators (to appear)},
        doi       = {10.1109/TPAMI.2023.3313648},
        year      = {2023},
        publisher = {IEEE Computer Society},
    }
\end{verbatim}

\vspace{5ex}

\noindent Please find \textbf{reference implementation} for t-EER computation, with examples:
\begin{itemize}
    \item \textbf{Google Colab}\\ \texttt{\url{https://colab.research.google.com/drive/1ga7eiKFP11wOFMuZjThLJlkBcwEG6_4m?usp=sharing}}\vspace{1ex}
    \item \textbf{Github:}\\
    \texttt{\url{https://github.com/TakHemlata/T-EER}}
\end{itemize}

\vspace{60ex}
\noindent \copyright 2023 IEEE. Personal use of this material is permitted. Permission
from IEEE must be obtained for all other uses, in any current or future
media, including reprinting/republishing this material for advertising or
promotional purposes, creating new collective works, for resale or
redistribution to servers or lists, or reuse of any copyrighted
component of this work in other works.

\newpage
\twocolumn

%
\title{t-EER: Parameter-Free Tandem Evaluation of Countermeasures and Biometric Comparators}
%
%
%
%

\author{Tomi~H.~Kinnunen, 
        Kong~Aik~Lee,~\IEEEmembership{Senior Member,~IEEE,}
        Hemlata~Tak, \\
        Nicholas~Evans,~\IEEEmembership{Member,~IEEE,}
        Andreas~Nautsch
        ~\IEEEmembership{}
\IEEEcompsocitemizethanks{\IEEEcompsocthanksitem T. H. Kinnunen is with the School of Computing, University of Eastern Finland (UEF), Finland, e-mail: tomi.kinnunen@uef.fi.\protect\\
\IEEEcompsocthanksitem K. A. Lee is with Singapore Institute of Technology, Singapore, and the Institute for Infocomm Research, A$^\star$ STAR, Singapore, e-mail: kongaik.lee@singaporetech.edu.sg.\protect\\
\IEEEcompsocthanksitem H. Tak and N. Evans are with EURECOM, France, e-mail: \{tak,evans\}@eurecom.fr.\protect\\
\IEEEcompsocthanksitem A. Nautsch was with Avignon Universit\'e, France (parts of this work were conceived while he was at EURECOM); he is now with goSmart GmbH, Germany.}
\thanks{Manuscript submitted to IEEE-T-PAMI 27/01/2023}}

%
%

\markboth{Journal of \LaTeX\ Class Files,~Vol.~14, No.~8, August~2015}%
{Shell \MakeLowercase{\textit{et al.}}: Bare Demo of IEEEtran.cls for Computer Society Journals}
%



\IEEEtitleabstractindextext{%
\begin{abstract}
Presentation attack (spoofing) detection (PAD) typically operates alongside biometric verification to improve reliablity in the face of spoofing attacks. Even though the two sub-systems operate in tandem to solve the single task of reliable biometric verification, they address different detection tasks and are hence typically evaluated separately. Evidence shows that this approach is suboptimal. We introduce a new metric for the joint evaluation of PAD solutions operating in situ with biometric verification. In contrast to the tandem detection cost function proposed recently, the new tandem equal error rate (t-EER) is parameter free. The combination of two classifiers nonetheless leads to a \emph{set} of operating points at which false alarm and miss rates are equal and also dependent upon the prevalence of attacks. We therefore introduce the \emph{concurrent} t-EER, a unique operating point which is invariable to the prevalence of attacks. Using both modality (and even application) agnostic simulated scores, as well as real scores for a voice biometrics application, we demonstrate application of the t-EER to a wide range of biometric system evaluations under attack. The proposed approach is a strong candidate metric for the tandem evaluation of PAD systems and biometric comparators. 
\end{abstract}

}

\maketitle

\IEEEdisplaynontitleabstractindextext

%
\IEEEpeerreviewmaketitle

\IEEEraisesectionheading{\section{Introduction}\label{sec:introduction}}

%
%
%
%

\IEEEPARstart{B}{iometric} recognition is nowadays in widespread use across forensic, civilian and consumer domains. Likewise,
approaches for the testing and performance reporting of biometric systems~\cite{Ruud2004-biometrics-guide,wayman2005biometric} under \emph{normal presentation mode}\footnote{\emph{Normal} or \emph{routine} implies that the system is used in the fashion intended by the system designer~\cite{ISOpresentationAtack}. Spoofed trials are considered to be outside of the normal presentation mode.} are well established --- see e.g.\ the ISO/IEC 19795 standard~\cite{ISO-IEC-197951-FDIS-2021}. 
Despite high reliability, biometric systems are unfortunately not infallible outside of the normal presentation mode, e.g.\ when they are attacked by an adversary. 
Since the identification of biometric system attack points more than two decades ago~\cite{Ratha2001-enhancing}, the community has been active in addressing vulnerabilities, especially \emph{presentation} or \emph{spoofing attacks}, for all the major biometric modes such as fingerprints~\cite{Galbally2019-fingerprint-PAD}, face~\cite{HernandezOrtega2019-face-PAD}, and voice~\cite{WU2015-spoofing-cm-survey,Liu2022ASVspoof2T}. Well-studied examples of spoofing attacks include printed photographs (face), gummy fingers (fingerprints), and audio replay (voice). Of particular concern are vulnerabilities to  \emph{DeepFakes}~\cite{Verdoliva2020-mediaforensics-and-deepfakes}, stemming from rapid developments in deep learning~\cite{Goodfellow2014-GAN}, which can be used to implement face swapping and voice cloning. 

Having recognized the threat, the ISO/IEC Joint Technical Committee (JTC) 1/SC 37 launched a multi-part 30107 standard series~\cite{ISOpresentationAtack} in 2016 to provide a foundation for what became known as \emph{presentation attack detection} (PAD), with a second version being released in 2023~\cite{ISO-IEC-30107-3:2023}.
As defined by the standard, attacks are assumed to be \emph{presented} at the sensor level of the targeted biometric data capture subsystem (such as a fingerprint reader, a camera, or a microphone).\footnote{In the authors' view, this definition excludes certain forms of attack that are relevant to specific biometric modes, particularly when the sensor is beyond the control of the biometric system designer.  In the particular case of speaker verification technology deployed in many call centers, users choose their \emph{own} biometric capture subsystem --- say their smartphone. Thus, besides presentation attacks directed at the microphone, an adversary may also inject attacks \emph{post-sensor} using, e.g.\ a voice modification system operating somewhere between the microphone and the call center. The ASVspoof initiative~\cite{Liu2022ASVspoof2T} differentiates the two cases, termed as \emph{physical attacks} (presented at the sensor) and \emph{logical attacks} (injected post-sensor).} PAD has since attracted considerable interest; attacks have potential to interfere with the intended operation of the biometric system and to substantially compromise performance unless prepared for. 

\begin{figure*}[!t]
    \centering
    \begin{subfigure}[b]{0.6\textwidth}
        \includegraphics[width=\textwidth]{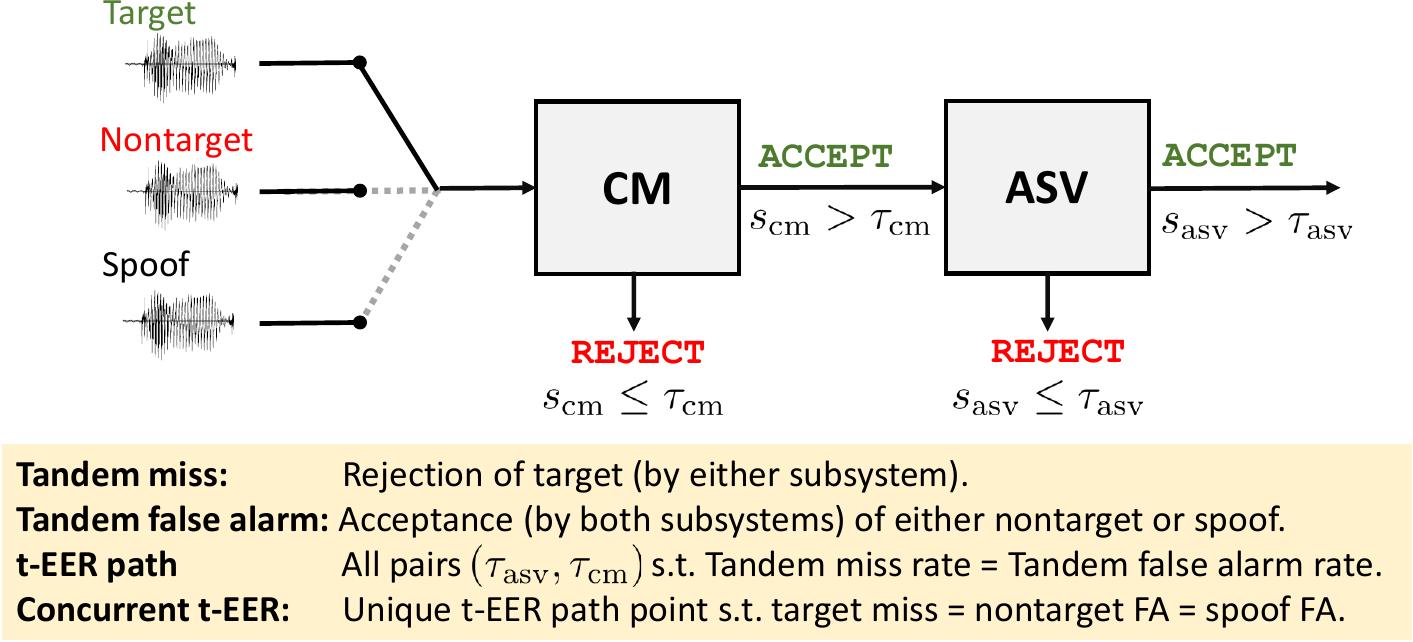}
        \caption{A tandem detection system and key definitions.}
        \label{subfig:tandem-system-diagram}
    \end{subfigure}
    \vspace{1ex}
    \begin{subfigure}[b]{0.47\textwidth}
        \centering
        \begin{tikzpicture}[scale=0.7,font=\footnotesize]
        \draw [->] (0,-2) -- (0, 2) node[above] {Score: CM};
        \draw [->] (-2,0) -- (2, 0) node[right] {Score: ASV};
        
        \draw[color=BrBG-4-3, pattern={Lines[angle=-45,distance={3pt/sqrt(2)}]}, pattern color=BrBG-4-3] (1, 1) rectangle node [black] {$A_1$} +(1,1);
        \draw[color=BrBG-4-2, pattern={Lines[angle=0,distance={3pt/sqrt(2)}]}, pattern color=BrBG-4-2] (1, 1) rectangle node [black] {$A_2$} +(-3,1);
        \draw[color=PiYG-4-3, pattern={Lines[angle=90,distance={3pt/sqrt(2)}]}, pattern color=PiYG-4-3] (1, 1) rectangle node [black] {$A_3$} +(1,-3);
        \draw[color=PiYG-4-2, pattern={Lines[angle=0,distance={3pt/sqrt(2)}]}, pattern color=PiYG-4-2] (1, 1) rectangle +(-3,-3);
        \draw[color=PiYG-4-2, pattern={Lines[angle=90,distance={3pt/sqrt(2)}]}, pattern color=PiYG-4-2] (1, 1) rectangle node [black] {$A_4$} +(-3,-3);
        
        \draw [densely dotted,ultra thick] (1,-2) node [below] {Second threshold: ASV} -- (1, 2);
        \draw [densely dashed,ultra thick] (-2, 1) node [left,xshift=-1em] {First threshold: CM} -- (2, 1);
        
        \draw[<->,yshift=-2em] (1,-2) -- node[midway,below,align=center] {$P_\text{fa,non}^\text{asv}$\\[1ex] $P_\text{fa,spoof}^\text{asv}$} (2,-2);
        \draw[<->,yshift=-2em] (-2,-2) -- node[midway,below] {$P_\text{miss}^\text{asv}$} (1,-2);
        \draw[<->,xshift=-.5em] (-2, 2) -- node[midway,left] {$P_\text{fa}^\text{cm}$} (-2, 1);
        \draw[<->,xshift=-.5em] (-2, 1) -- node[midway,left] {$P_\text{miss}^\text{cm}$} (-2, -2);
    \end{tikzpicture}
        \caption{The 2D space of the detection scores $(s_\text{asv},s_\text{cm})$ and thresholds $\vec{\tau}=(\tau_\text{asv},\tau_\text{cm})$.}
        \label{subfig:areas}
    \end{subfigure}
    \hfill
    \begin{subfigure}[b]{0.47\textwidth}
        \centering
        \includegraphics[width=\textwidth]{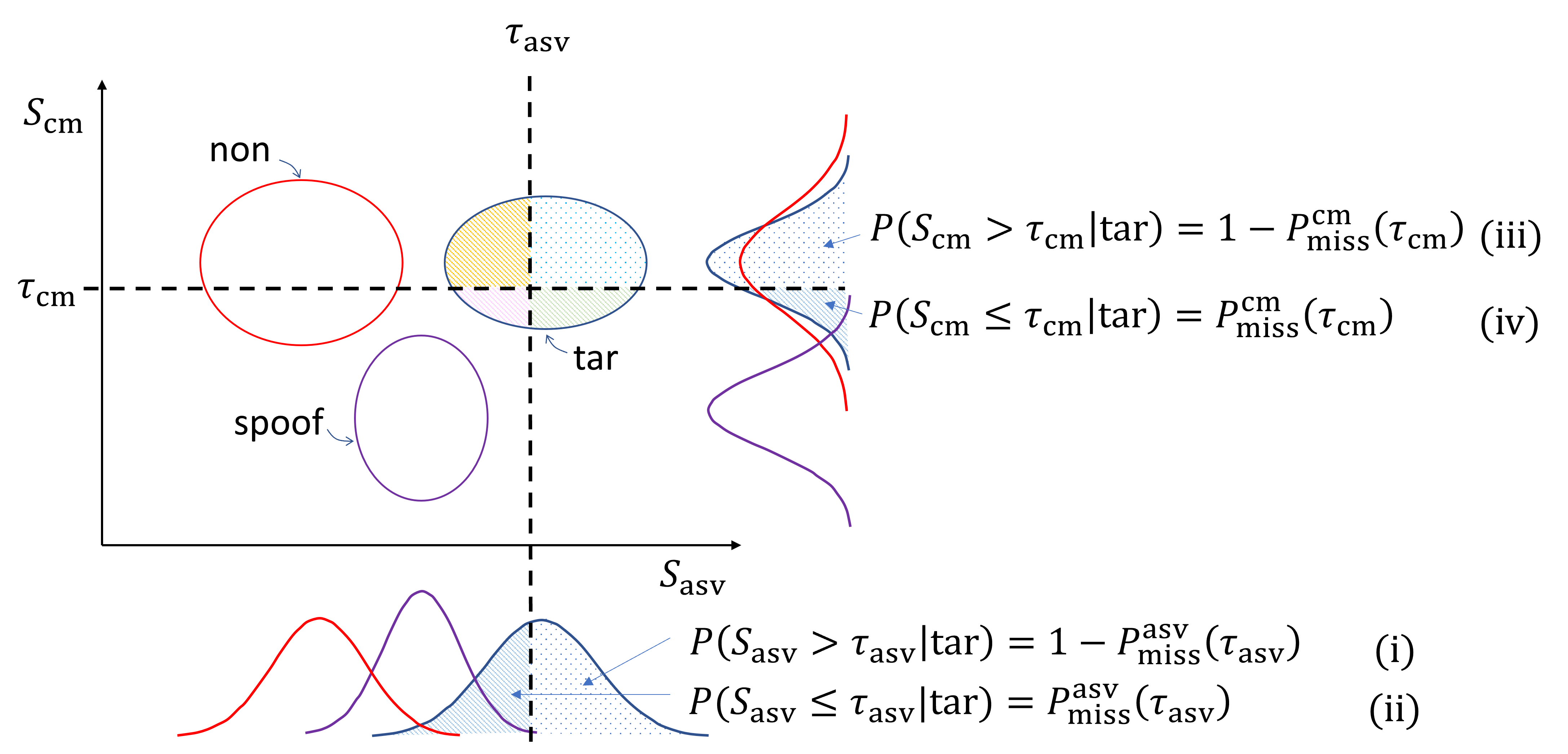}
        \caption{Similar 2D score space as in (b), with computation of per-system error probabilities visualized.}
        \label{subfig:missfa}
    \end{subfigure}
    \caption{The \emph{tandem equal error rate} (t-EER) is a metric for the evaluation of biometric systems comprising a cascaded spoofing countermeasure (CM) and comparator such as an automatic speaker verification (ASV) system, illustrated in \textbf{(a)}. Extending the conventional EER, the t-EER is found by adjusting the \emph{two} thresholds \textbf{(b)} to equalise total miss and false alarm rates. This leads to a \emph{function} with multiple possible t-EERs, summarized using a unique scalar, the \emph{concurrent} t-EER. As illustrated in \textbf{(b)}, in the tandem score space, we compose two linear decision boundaries for scores of the two subsystems. In \textbf{(c)}, the miss rate of the target class corresponds to the shaded area, which includes trials with both ASV and CM scores fall below the two thresholds, $\tau_\text{asv}$ and $\tau_\text{cm}$. The marginal distributions shown are those of ASV system (below) and the CM (right).}.
    \label{fig:combined}
\end{figure*}
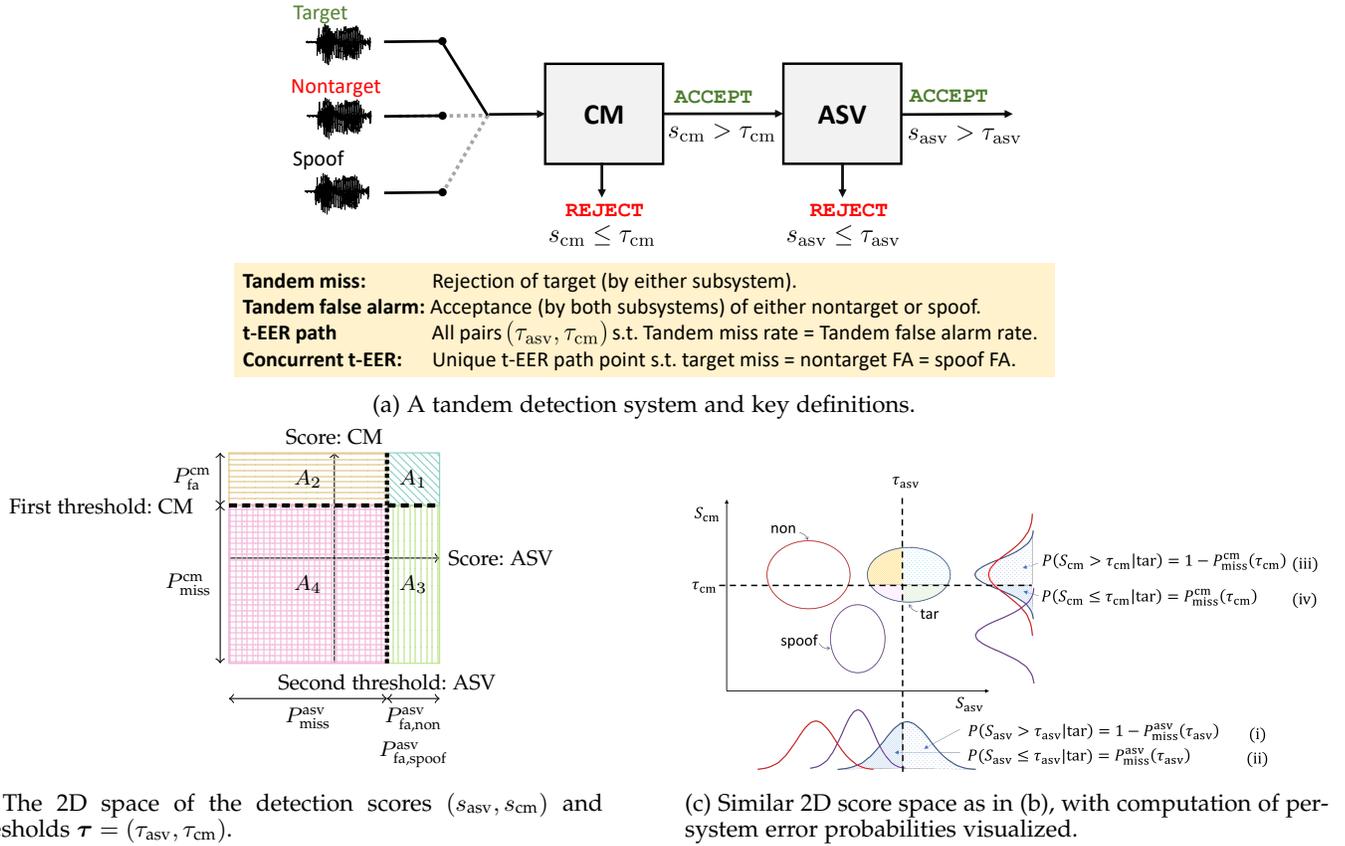

PAD is today an established research topic within the biometrics research community and one that is still attracting growing attention. While the practical solutions vary from challenge-response approaches~\cite{HernandezOrtega2019-face-PAD} to liveness detection~\cite{Micheletto2022-Livdet-review}, quality assessment~\cite{Galbally2019-fingerprint-PAD}, and artifact detection~\cite{wang2021comparative} specific to each biometric mode, most take the form of binary classifiers whose role is to determine whether or not a given biometric probe is a bona fide sample (real) or a presentation attack (fake).

Whatever the approach, the PAD problem has been treated historically as a separate task that is independent of the key biometric comparison task. The latter consists in verifying whether or not a pair of biometric samples in the form of reference (enrolment) and probe (test) samples originate from the same biometric capture subject (individual).
As illustrated in Fig.~\ref{subfig:tandem-system-diagram}, 
practical biometric authentication solutions typically combine separate PAD and biometric comparator sub-systems operating in tandem.

\subsection{The Need for Tandem Assessment} 

Both sub-systems are binary classifiers, each designed to solve a given, different task. Nonetheless, they are combined to solve what is, after all, the \emph{single} task of reliable biometric verification. \textbf{We argue that, no matter what the architecture used to combine them, the two sub-systems should be evaluated as a whole}. As noted in \cite{Chingovska2019}, however, the PAD subsystem and the biometric comparator are typically evaluated separately. 

We suspect there are two main reasons for this. The first is one of convenience which follows the well-known \emph{Maslow's hammer} law, \emph{`if the only tool you have is a hammer, it is tempting to treat everything as if it were a nail`}; since both the PAD subsystem and biometric comparator are binary classifiers, any established performance assessment tool and figure of merit ~\cite{wayman2005biometric,DODDINGTON2000-NIST} are readily applicable.\footnote{With the understanding that the positive and negative classes and error measures could be termed differently for clarity; for instance, in biometric verification the proportion of negative class cases that were falsely accepted is known as \emph{false acceptance rate} (FAR), \emph{false match rate} (FMR), or \emph{false alarm rate} depending on the context. In in ISO/IEC 30107 \cite{ISOpresentationAtack}, the equivalent property of a PAD subsystem is known as \emph{attack presentation classification error rate} (APCER).} While the separate assessment of the two subsystems is useful for analysis purposes, it effectively forces a \emph{ternary} classification task (bonafide target, nontarget, spoofing attack) to be treated as two separate binary tasks~\cite{Chingovska2014-biometrics-eval-under-spoofing-attacks}. This approach not only \emph{increases} the amount of numbers to be reported (e.g.\ a pair of EERs corresponding to the PAD subsystem and the biometric comparator) but, worse, provides no guarantees on the performance of the complete biometric verification system as a whole. Previous work~\cite{Kinnunen2020-tandem-fundamentals} confirms that the best-performing PAD subsystem as judged from separate approaches to assessment does not always offer the best protection to the biometric comparator.  The separate assessment of the two subsystems is hence sub-optimal.

This leads us to the second suspected reason --- even if one might be convinced that the two systems should be evaluated jointly, there are no commonly agreed guidelines on \emph{how} this should be done. While score- and decision-level combination~\cite{Marasco2012-liveness-fingerprint-combination,Sahid2016-integrated,Korshunov2017-scorefusion,Todisco2018-integrated} are popular choices, they can be implemented using various different approaches, with possible tradeoffs~\cite{Chingovska2013-joint-operation} in the treatment of bonafide users and spoofing attacks. In fact, even the ISO/IEC 30107 standard leaves the details of the PAD system integration to the general biometric framework open-ended, by noting that the PAD subsystem could be placed e.g.\ before or after the biometric comparator subsystem, and/or at different levels. The standard explicitly acknowledges that \emph{``Outside the scope are \dots overall system-level security or vulnerability assessment''}. While one can expect carefully-tailored solutions for a given biometric mode, the fusion of multiple biometric modes, or the fusion of multiple PAD subsystems~\cite{Sahid2016-integrated} to perform well, such solutions tend to be complex and specific to the selected application. \textbf{There is clear scope for a common reference approach to the tandem assessment of arbitrary biometric comparator and PAD subsystems.} 

\subsection{Our Novelty: the Tandem Equal Error Rate (t-EER)}

The authors recently proposed a new decision-theoretic framework, namely the
\emph{tandem detection cost function} (t-DCF) \cite{Kinnunen2020-tandem-fundamentals}, as a solution to the performance testing of cascaded biometric comparator and PAD subsystems as illustrated in Fig.~\ref{subfig:tandem-system-diagram}. The t-DCF was developed as a two-system, three-class generalization of the widely-adopted \emph{detection cost function} (DCF)~\cite{DODDINGTON2000-NIST} endorsed by the National Institute of Standards and Technology (NIST). It was adopted as the primary evaluation metric for the two most recent ASVspoof challenge editions~\cite{Liu2022ASVspoof2T}, and has now been broadly adopted by the voice PAD and biometrics community, even if not embraced by other biometrics communities. One detraction of both the DCF and the t-DCF is that they require the choosing of two sets of parameters --- detection costs and class priors --- which can be difficult in practice, or feel arbitrary. While the authors generally advocate the adoption of performance measures that make the uncertainty of class prevalence (priors) and implications of decisions (costs) explicit, we recognise the appeal of parameter-free, more easily interpretable metrics. Remaining convinced in the importance of tandem evaluation, we have hence set out to derive an alternative to the t-DCF that is independent of costs and priors.

Our starting point is the traditional \emph{equal error rate} (EER), a popular metric used for the assessment of any ordinary binary classifier. The EER is defined as the error rate at a particular, special operating point at which the false alarm and miss rates are equal. It is also independent of priors, cost parameters and requires no \emph{a priori} setting of detection thresholds~\cite{Kinnunen2020-tandem-fundamentals}. 
Despite acknowledged shortcomings~\cite{ISO-IEC-197951-FDIS-2021,ISOpresentationAtack}, the EER remains among the most widely-adopted metrics in biometrics research and is used broadly for optimisation and model selection. Since the EER also serves as an upper bound to the Bayes error rate~\cite{BrummerFS21}, it is often also used as an empirical target in preliminary investigations of new features or model architectures. 

We report in this paper an extension to the conventional EER which supports the evaluation of a pair of binary classifiers operating in tandem. The new evaluation methodology is broadly applicable to any such scenario, even those beyond the study of biometrics.  
The specific application domain upon which we focus upon in this paper involves the study and evaluation of combined biometric comparators and PAD subsystems. The proposed common reference approach to their tandem evaluation remains agnostic to any specific biometric (e.g.\ face, iris and fingerprint). To highlight its broad application to both non-biometric and biometric scenarios alike, we report experiments with sets of simulated scores. Nonetheless, in order to demonstrate an application of the new evaluation methodology in a real, practical scenario, we also report experimental work involving voice biometrics. 

We summarize the organization, as well as key novelties of the paper, as follows.
\begin{itemize}
    \item After setting out the detection-theoretic foundations for the evaluation of classical standalone classifiers in Section~\ref{sec:preliminaries}, Section~\ref{sec:tandem-systems} provides a self-contained formulation of the tandem detection task; 
    \item The tandem equal error rate (t-EER), and the extension of a classical operating \emph{point} to an operating \emph{path}, as detailed in Section~\ref{sec:tandem-equal-error-rate}, are new concepts not presented in any prior work. As we will see, the t-EER can be viewed as a \emph{unifying} metric from which the classical per-system (PAD and biometric comparator) EERs are obtained as special cases;
    \item Section~\ref{sec:teer-computation} describes a practical and efficient approach to compute the t-EER path. For practitioners, we provide an open-source reference implementation;
    \item Unlike the classical scalar-valued EER, the new t-EER represents a \emph{function} with the t-EER path as its domain. Section~\ref{sec:concurrent-teer} shows how to summarize that path as a unique scalar in an intuitive, meaningful way. The result, coined the \emph{concurrent t-EER}, corresponds to a unique pair of thresholds at which the tandem system miss rate, the false alarm rate for nontargets, and the false alarm rate for spoofing attacks are all equal. The proposed metric hence reflects equal treatment of all three classes.
    \item Section \ref{sec:experiments} demonstrates use of the proposed metric in the context of voice biometrics. Finally, conclusions are presented in Section~\ref{sec:conclusions}.
\end{itemize}

\section{Preliminaries}\label{sec:preliminaries}

The proposed metric evaluates a pair of PAD and biometric comparator as illustrated in Fig. \ref{subfig:tandem-system-diagram}. Before detailing the combination of the two systems in Section \ref{sec:tandem-systems}, we first introduce basic notation and describe the evaluation of single detectors as a starting point. For a broader introduction to the evaluation of biometric systems, the interested reader is pointed to \cite{Ruud2004-biometrics-guide,wayman2005biometric}. Evaluation of both biometric verification \cite{Ruud2004-biometrics-guide,wayman2005biometric} and PAD systems \cite{Chingovska2014-biometrics-eval-under-spoofing-attacks} is rooted in detection theory (e.g.   
\cite{ROC-curves-2009,Macmillan2005-detection-user-guide}).

\subsection{Detectors}

Let $(x,y)$ denote paired input $x \in \mathscr{X}$ (an element of some signal or feature space $\mathscr{X}$) and the corresponding ground-truth label $y \in \mathscr{Y} \mydef \{0,1\}$. A \emph{binary classifier}, or a \emph{detector}, is a parameterized\footnote{Here, $\vec{\theta} \in \Theta$ denotes model parameters, for instance, the set of weights and biases of a neural network.} function $g_{\vec{\theta}}: \mathscr{X} \rightarrow \mathscr{Y}$ which makes a prediction, $\hat{y}$, of the label of $x$. It does so in two steps. First, the detector computes a real-valued \emph{score}, $s(x) \mydef s \in \mathbb{R}$ with the convention that higher numerical values are associated with greater confidence for $x$ to originate from the positive class ($y=1$). Typically $s$ is a likelihood ratio, log-likelihood ratio, or class posterior produced by a model (such as a deep neural network). In the second stage, the predicted label $\hat{y}$ (that may differ from the ground-truth label $y$) is obtained by comparing the score to a preset threshold, $\tau$:
\begin{equation}\label{eq:score-to-decision}
    \hat{y}=\left\{
        \begin{aligned}
            1 &, \;\;\; \text{if}\;\; s > \tau\\
            0 &, \;\;\; \text{if}\;\; s \leq \tau.
        \end{aligned}
    \right. 
\end{equation}

In biometric verification, $x$ consists of a paired representation $x=(x_\text{e},x_\text{t})$ of enrollment $(x_\text{e})$ and test $(x_\text{t})$ samples, also known as the \emph{biometric reference} and \emph{biometric probe}, respectively~\cite{iso-iec:2382-biometric-vocabulary}. Depending on the system, these samples could be represented by raw signal/pixel values or feature vectors computed from them. The ground truth label $y$, in turn, serves to indicate whether or not the enrolment and test samples belong to the same individual (\emph{a target}) or not (\emph{a nontarget}). The ISO/IEC 2382-37 standard~\cite{iso-iec:2382-biometric-vocabulary} refers instead to 
\emph{biometric mated/non-mated comparison trials}. For the second task considered in this study, PAD, $x$ is (usually) a representation of the test sample alone, whereas $y$ serves to indicate if $x$ is real (\emph{bona fide} human) or a fake (\emph{spoofed}) sample. The corresponding terminologies in the ISO/IEC 30107 standard~\cite{ISOpresentationAtack} 
are \emph{bona fide / attack presentations}.

The approach to evaluation and the metric reported in this paper are agnostic to the semantics of $x$ and the classifier $g_{\vec\theta}$. The only data available to us, as evaluators, are the finite set of detection scores and the ground-truth labels, $\{(s_i,y_i): i=1,\dots,N\}$  for a test set of $N$ samples. From here on we drop the instance subscript $i$ and treat both the scores and their labels as realizations of random variables $S$ and $Y$, respectively. To this end, we use $F_{S|Y}(\tau|Y) \mydef P(S \leq \tau|Y)$ to denote the cumulative density function (CDF) of scores for class $Y$. Here, $P(A|B)$ is the standard notation of conditional probability --- the probability of event $A$ given that event $B$ has occured.

\subsection{Detection error rates}

Any binary classifier is prone to two different types of detection errors --- \emph{misses} (false rejections) and \emph{false alarms} (false acceptances). The former take place when the true class of $x$ is 1 but the classifier predicts 0 (rejects $x$). The latter happen when the true class is 0 but the classifier predicts 1 (accepts $x$). These two error rates arise naturally from the \emph{total error rate}, $P_\mathscr{E}$, defined as the probability that the predicted label differs from the true label. Following the rules of joint and conditional probabilities \cite{bishop2006}, 
    \begin{equation}
        \begin{aligned}
            P_\mathscr{E} & \mydef P(\hat{Y} \neq Y)\\
            & = P(\hat{Y}{=}0, Y{=}1) + P(\hat{Y}{=}1, Y{=}0)\\
            & = P(Y{=}1)P(\hat{Y}{=}0|Y{=}1) + P(Y{=}0)P(\hat{Y}{=}1|Y{=}0)\label{eq:total-error-rate}\\
            & = \DBprior P_\text{miss} +  (1 - \DBprior)P_\text{fa},
        \end{aligned}
    \end{equation}
where
\begin{equation}
    \begin{aligned}
    P_\text{miss} & \mydef P(\hat{Y}=0|Y=1)\\
    P_\text{fa} & \mydef P(\hat{Y}=1|Y=0)
\label{def:miss-fa-rates}
\end{aligned}
\end{equation}
are the miss and false alarm rates and
    \begin{equation}\label{eq:dbprior-single-system}
        \DBprior \mydef P(Y=1)         
    \end{equation}
denote a \emph{database prior}. As the name suggests, it reflects the relative proportion of positive and negative class samples in the database protocol, known by the evaluator. The database prior is \emph{not} to be confused with a user-supplied \emph{asserted} prior one might have about future data in a predictive context; whereas the database prior is a deterministic property of a database protocol, an asserted prior reflects one's uncertainty of class priors in a future environment. In contrast to the DCF \cite{DODDINGTON2000-NIST} and the t-DCF \cite{Kinnunen2020-tandem-fundamentals} metrics that require asserted priors, the concurrent t-EER metric proposed in this study has no dependencies, neither database nor asserted priors. Note that if the database contains an equal proportion of trials from the two classes, the database prior is $\DBprior=\frac{1}{2}$. If the total error rate were to be used as a performance measure, the relative class proportions impose implicit weighting of the two error types given by the convex combination in \eqref{eq:total-error-rate}. Such dependency on database class proportions is undesirable, not least because the error rates are noncomparable across datasets with different class proportions.

By combining \eqref{eq:score-to-decision} with \eqref{def:miss-fa-rates}, the miss rate can be expressed as a function of $\tau$, as
    \begin{equation}
        \begin{aligned}
            P_\text{miss}(\tau) & = P(\hat{Y}=0|Y=1)\\
            & = P(S \leq \tau |Y=1)\\
            & = F_{S|Y=1}(\tau|Y=1).
        \end{aligned}
    \end{equation}
Similarly, the false alarm rate can be expressed as
    \begin{equation}
        \begin{aligned}
            P_\text{fa}(\tau) & = P(\hat{Y}=1|Y=0)\\
            & = P(S > \tau | Y=0)\\
            & = 1 - F_{S|Y=0}(\tau|Y=0).
        \end{aligned}
    \end{equation}

A central notion in \emph{receiver operating characteristic} (ROC) \cite{ROC-curves-2009,Macmillan2005-detection-user-guide} and \emph{detection error trade-off} (DET) \cite{martindet} analysis is to observe the trade off between miss and false alarm rates by varying $\tau$. In particular, the miss and the false alarm rates are, respectively, non-decreasing and non-increasing functions of $\tau$,
    \begin{equation}
        \begin{aligned}
            P_\text{miss}(\tau) & \geq P_\text{miss}(\tau') & \text{whenever} \;\; \tau \geq \tau'   \nonumber\\
            P_\text{fa}(\tau) & \leq P_\text{fa}(\tau') & \text{whenever} \;\; \tau \geq \tau',   
        \end{aligned}
    \end{equation}
with the following limits which follow directly from the properties of the CDF: 
    \begin{equation}
        \begin{aligned}
        \lim_{\tau \rightarrow -\infty} P_\text{miss}(\tau) & =0, & \;\;\;\lim_{\tau \rightarrow +\infty} P_\text{miss}(\tau)& =1,\nonumber\\
        \lim_{\tau \rightarrow -\infty} P_\text{fa}(\tau)& =1, & \;\;\;\lim_{\tau \rightarrow +\infty} P_\text{fa}(\tau)& =0.\nonumber
        \end{aligned}
    \end{equation}
In the remainder we drop `lim' and refer to these limits simply by setting $\tau=-\infty$ or $\tau=\infty$. In biometric verification and spoofing detection, $\tau$ can be used to control the trade-off between user convenience and security. The `accept all' setting $\tau=-\infty$ leads to all inputs being accepted, giving $P_\text{miss}=0$ and $P_\text{fa}=1$. Likewise, $\tau=+\infty$ leads to a `reject all' system, for which $P_\text{miss}=1$ and $P_\text{miss}=0$. While these extreme cases are referred to in the analyses which follow, for any practical detection system, $|\tau| < \infty$.

\subsection{Error rate tradeoff and performance summarization}

By varying $\tau$, one obtains a ROC/DET curve which displays one error rate as a function of another. ROC curves \cite{ROC-curves-2009,Fawcett2005-an-intro-to-ROC-analysis} are obtained by plotting $P_\text{fa}(\tau)$ against $1-P_\text{miss}(\tau)$. DET curves \cite{martindet}, used by default in voice biometrics research (and also recommended for PAD evaluation \cite{ISOpresentationAtack}), graph $P_\text{fa}(\tau)$ against $P_\text{miss}(\tau)$ on a normal deviate designed to highlight better the differences between high-performing systems. The interested reader is pointed to \cite[Section 4]{Nautsch2019-PhDthesis} for a review and additional approaches to plot error trade-offs.

For performance assessment, a single scalar is preferred over a function. Since no scalar can preserve all the details of the full error tradeoff function, numerous performance indices have been proposed to summarize the aspects relevant in a given domain (for reviews see e.g.\ \cite{Hand2021-assessing-performance,Tharwat2020-classification-assessment}). In biometric verification and PAD research papers, the EER (defined below) is among the most popular choice. Others include the \emph{area under the curve} (AUC)~\cite{Fawcett2005-an-intro-to-ROC-analysis}, the $F_1$ score and the \emph{cost of log-likelihood ratio} ($C_\text{llr}$) \cite{BRUMMER2006-application-independent}. In contrast to these non-parametric indices, \emph{weighted} error metrics include the \emph{detection cost function} (DCF) \cite{DODDINGTON2000-NIST} which measures the average (expected) cost of making an error. Beyond biometrics, readers are referred to \cite{Hand2021-notes-on-hmeasure} for an interesting measure which imposes a weight function (prior distribution) over the detection costs. 

For any operational automated biometric verification system, one must set the detection threshold $\tau$ in advance to obtain a desirable trade-off between security and convenience as is normally set according to knowledge of the operational environment. This can be done by optimizing a criterion of interest (such as the EER or DCF) using development data or by fixing the threshold according to the analytic form provided by Bayes' minimum-risk threshold. Since the scores produced by a model --- even by a probabilistic model --- rarely present calibrated probabilities, one may combine this approach with a \emph{score calibration} model \cite{BRUMMER2006-application-independent,Ferrer2022-robust-calibration} trained using development data. Graphical presentations which emphasize this aspect of evaluation --- improving the generality of threshold setting or score calibration from development to evaluation data --- include the \emph{expected performance curve} (EPC) \cite{bengio04b-EPC} and the \emph{applied probability of error} (APE) curve \cite{BRUMMER2006-application-independent}. The EPC was further generalized in \cite{Chingovska2014-biometrics-eval-under-spoofing-attacks} to an \emph{expected performance and spoofability curve} (EPS) for the evaluation of biometric systems under attack. Since our focus is on a \emph{discriminative} criterion (EER) rather than calibration, threshold selection \footnote{For \emph{known} class-conditional feature distributions and class priors, the \emph{optimal} detection score for a detection task is the log-likelihood ratio (LLR) of the null and alternative hypothesis likelihoods, leading to optimal (lowest error rate) decision threshold $\tau_\text{Bayes} := -\text{logit}(\pi)$, where $\pi$ denotes the prior of the target (positive) class and $\text{logit}(\pi) := \log(\pi) - \log(1-\pi)$. In any practical setting, however, neither the class priors nor the class-conditional data distributions are known, but have to be estimated from data using a model. This results in \emph{miscalibrated} scores that no longer produce optimal decisions derived using $\tau_\text{Bayes}$. The EER, however, measures the degree of overlap between the score distributions of each class, ignoring completely the scale of the detection scores. Just as the entire ROC or DET curve, it is agnostic to any global order-preserving score transforms (such as the affine map $s \mapsto as + b$ with $a>0$).}
and score calibration is outside the scope of this paper. As we will see, the generalisation of the EER to the evaluation of two-systems and three classes is not straightforward. Before doing so, we present the definition of the classical EER.

\subsection{Equal error rate (EER)}\label{subsec:classic-EER}

Noting the monotonicity and contrasting values of miss and false alarm rates at $\pm \infty$, there always exists an interval $\Omega \subset \mathbb{R}$ of thresholds in which the miss and the false alarm rates are equal\footnote{Technically, this is guaranteed only when the actual score CDFs are continuous. In practical scenarios, empirical CDFs are constructed from finite sets (typically by error counting) and there may not be an exact threshold where the two are equal. While in this study we rely on simple nearest-neigbor interpolation to determine the EER, the interested reader is pointed to \cite{Brummer2013-bosaris} for more advanced methods based on the ROC convex hull \cite[Section 6]{Fawcett2005-an-intro-to-ROC-analysis} to determine the EER.}: $P_\text{miss}(\tau)=P_\text{fa}(\tau)$, for all $\tau \in \Omega$. Usually this interval collapses to a single point (a scalar). We define the \emph{equal error rate} (EER) as the value $\text{EER} \mydef P_\text{miss}(\tau)=P_\text{fa}(\tau)$ where $\tau \in \Omega$ is the interval at which the miss and false alarm rates have the same value. Apart from some special cases, such as as Gaussian score PDFs with known parameters \cite{Poh2004-analytic-EER,leeuwen13_interspeech}, neither the EER nor the EER threshold have closed-form solutions, hence they are determined empirically from the operating point at which the empirical miss and false alarm functions intersect.

The EER is closely related to the total error rate; in fact, substituting  $P_\text{miss}=P_\text{fa}=\text{EER}$ into \eqref{eq:total-error-rate} shows they equal one another at the EER operating point. Further, as detailed in \cite{BrummerFS21}, under calibrated probabilities, the EER can be interpreted as a strict upper bound to the Bayes error rate: 
    \begin{equation}\label{eq:EER-as-max-total-error-rate}
        \max_{\pi} P_\mathscr{E} = \text{EER}.
    \end{equation}
Informally, the reader may verify \eqref{eq:EER-as-max-total-error-rate} by differentiating \eqref{eq:total-error-rate} w.r.t. $\pi$ which yields the EER condition $P_\text{miss}=P_\text{fa}$. Hence, when used as an empirical model selection criterion, the EER can be interpreted as the total error rate with the \emph{worst-possible} prior.

\section{Tandem systems}\label{sec:tandem-systems}

Consider now the case of a combined system comprising \emph{two} binary classifiers operating in tandem (here, PAD and biometric comparator subsystems). Each classifier addresses a different detection task. We want to combine their predictions in a manner which supports their tandem evaluation. As above, each of these two classifiers has two potential prediction outcomes --- one positive and one negative. The combined (tandem) system acts as a logical \texttt{AND} gate by producing an overall positive decision if (and only if) both classifiers produce positive predictions. In the remainder of this paper, we adopt special terminology used in the context of \emph{automatic speaker verification} (ASV) --- with the understanding that `ASV' can be replaced by `fingerprint', `face', `iris' verification (or any other biometric mode) without changing the methodology.

\subsection{Database Prior}

We now consider two sets of labels, $Y_\text{cm}=\{\text{bona fide}=1,\text{spoof}=0\}$ (label set of a countermeasure) and $Y_\text{asv}=\{\text{target}=1,\text{nontarget}=0\}$ (label set of an ASV system). We define the set of \emph{tandem labels}, $Y_\text{tdm}$, as the cartesian product $Y_\text{tdm} =Y_\text{cm} \times Y_\text{asv}$ which consists of four possible elements. 
The relative frequency of each of the four cases is given by the joint probability distribution over the two sets of ground-truth labels. Using the shorthand notation $P(Y_\text{cm}=y_\text{cm},Y_\text{asv}=y_\text{asv})=P(y_\text{cm},y_\text{asv})$ to drop random variable names, the joint probability distribution for the four possible tandem labels are given by
    \begin{equation}\label{eq:prior-joint-distribution}
        \begin{aligned}
            P(\text{bona},\text{tar}) & = P(\text{bona}|\text{tar})P(\text{tar})\\ 
            P(\text{bona},\text{non}) &  = P(\text{bona}|\text{non})P(\text{non})\\ 
            P(\text{spoof},\text{tar}) & = P(\text{tar}|\text{spoof})P(\text{spoof})\\ 
            P(\text{spoof},\text{non})& = P(\text{non}|\text{spoof})P(\text{spoof}),\\ 
        \end{aligned}
    \end{equation}
where class names are shortened for brevity. Given the specific tandem tasks of spoofing and speaker detection, we now assert three \emph{necessary} assumptions relevant to the conditional probabilities in the right-hand side of \eqref{eq:prior-joint-distribution}:
    \begin{equation}
        \begin{aligned}
            \textbf{(A1.)}\;\; & P(\text{bona}|\text{tar}) =1 & \text{Target speaker is bona fide.}\nonumber\\
            \textbf{(A2.)}\;\; & P(\text{bona}|\text{non}) =1 & \text{Zero-effort impostor is bona fide.}\\
            \textbf{(A3.)}\;\; & P(\text{tar}|\text{spoof}) =1 & \text{A spoofing attack is targeted.}\\
            \Big(\textbf{(A4.)}\;\; & P(\text{non}|\text{spoof}) =0 & \text{(\emph{follows from A3})}\Big)\\
        \end{aligned}
    \end{equation}
The first assumption (A1) means that users who claim their own identity present only bona fide utterances; the second (A2)  dictates that non-target or zero-effort impostor presentations contain bona fide speech (not spoofed speech). Finally, (A3) states that spoofed presentations are made to claim only a specific target (speaker) identity. These assumptions reflect a practical verification scenario in which spoofing attacks can be expected. Note that (A4) follows from (A3). Note also that while ISO/IEC 30107 \cite{ISOpresentationAtack} considers both subversive \emph{biometric impostors} (spoofing) and \emph{biometric concealers} (evasion), our focus is on the former.

By substituting assumptions (A1)--(A3) into \eqref{eq:prior-joint-distribution} we arrive at
    \begin{equation}\label{eq:tandem-prior}
        \begin{aligned}
            P(\text{bona},\text{tar})& =P(\text{tar}) & =: &\; \pi_\text{tar}\\
            P(\text{bona},\text{non})& =P(\text{non}) & =: &\; \pi_\text{non}\\
            P(\text{spoof},\text{tar})& =P(\text{spoof}) & =: &\; \pi_\text{spoof}\\
            P(\text{spoof},\text{non})& = 0, \\
        \end{aligned}
    \end{equation}
where $\pi_\text{tar}$, $\pi_\text{non}$ and $\pi_\text{spoof}$ are nonnegative and sum up to 1.
Because of these probabilistic constraints, the database prior $\vec{\pi}=(\pi_\text{tar},\pi_\text{non},\pi_\text{spoof})$ is now uniquely specified by two nonnegative numbers. The above prior definitions align with the notation in \cite{Kinnunen2020-tandem-fundamentals} though here we make the underlying assumptions explicit. From here on, we use `target', `nontarget' and `spoof' to refer to the tandem classes (bona, tar), (bona, non) and (spoof, tar), respectively. In the ISO/IEC 30107 standard \cite{ISOpresentationAtack}, the corresponding terminologies are \emph{bona fide mated} (target), \emph{bona fide non-mated} (nontarget), and \emph{attack} (spoof).

\subsection{Decision Rules}

The tandem detection system combines information extracted by the two subsystems --- CM and ASV --- to make the final accept/reject decision. This means that, in order for a particular identity claim to be accepted, the provided test utterance must be declared as bona fide \textbf{and} it must match the claimed identity in terms of speaker characteristics.

Depending on the outputs provided by the individual detectors, the combined decision could be obtained, for instance, by \texttt{AND}ing the hard decisions by 
\begin{equation}\label{eq:AND-combination}
    \begin{aligned}
        D & = \texttt{AND}\big( \mathbb{I}(s_\text{asv} > \tau_\text{asv}), \mathbb{I}(s_\text{cm} > \tau_\text{cm})\big)\\
        & = \mathbb{I}(s_\text{asv} > \tau_\text{asv})\cdot \mathbb{I}(s_\text{cm} > \tau_\text{cm})
    \end{aligned}
\end{equation}
where the indicator function $\mathbb{I}(\alpha)=1$ for a true proposition $\alpha$ (and 0 otherwise), and $D \in \{0,1\}$ stands for `decision': $D=0$ implies a negative prediction (reject) and $D=1$ a positive prediction (accept). The combined decision could alternatively be produced by passing the detection scores to a back-end classifier followed by thresholding:
    \begin{equation}\label{eq:fusion-combination}
        D = \mathbb{I}(\texttt{fuse}([s_\text{asv},s_\text{cm}]) > \tau_\text{fuse}),
    \end{equation}
where $\texttt{fuse}: \mathbb{R}^2 \rightarrow \mathbb{R}$ denotes an arbitrary back-end classifier with an associated single detection threshold, $\tau_\text{fuse}$. 

The theory developed in this study focuses on decision rules of the form \eqref{eq:AND-combination} where we assume access to two sets of detection scores and where the two detection thresholds $\tau_\text{asv}$ and $\tau_\text{cm}$ can vary concurrently. This leads to an extension of the  single-system detection theory where the concept of a scalar threshold is extended to a vector-valued decision boundary in a 2D score space.

While \eqref{eq:AND-combination} assumes statistical independence of the decisions produced by the two systems (see also~\cite[Section VIII.A]{Kinnunen2020-tandem-fundamentals}), we argue that this is reasonable whenever the systems address \emph{different} detection tasks -- as is the case here. Moreover, limiting our attention to 2-parameter decision rules specified by two orthogonal hyperplanes (see Fig.~\ref{subfig:areas}) enables tractable analysis (within a reduced space of combination rules). Note that the second, score fusion strategy \eqref{eq:fusion-combination} involves all possible classification rules that map continuous 2D vector onto a scalar. This includes, for instance, affine score combination, Gaussian back-end \cite{Todisco2018-integrated} and arbitrary feedforward DNNs with softmax output.

Another benefit of \eqref{eq:AND-combination} over the score fusion approach \eqref{eq:fusion-combination} is that we do not need \emph{paired} CM and ASV scores. Due to the statistical independence assumption of ASV and CM decisions, the error rate expressions of the tandem system (detailed in Section \ref{sec:tandem-equal-error-rate}) involves marginal (subsystem-specific) error rates and their products only. An important practical consequence is that the ASV and CM systems could be produced by two different parties that may not have the capacity or expertise to implement both. This property has been exploited in the organisation of the ASVspoof challenge series~\cite{Liu2022ASVspoof2T} for which an ASV system is implemented by the organizers, allowing participants to foucs upon the design of CMs.

\subsection{Miss and False Alarm Rates}

In summary, we assume three mutually exclusive classes (target, nontarget, spoof), and two possible decisions ($D=0$ or $D=1$) produced by the tandem system. Table \ref{tab:tandem-error-cases} indicates which cases consitute a classification error.
    \begin{table}[h]
        \centering
        \caption{Cross-tabulation of tandem system decisions (rows) vs. ground-truth labels (columns) encodes which cases consitute a classification error.}
        \begin{tabular}{|c|ccc|}
            \hline
            & Target & Nontarget & Spoof\\
            \hline\hline                         
                 $D=0$ & \textcolor{red}{\textbf{error}} & \textcolor{DarkGreen}{\textbf{OK}} & \textcolor{DarkGreen}{\textbf{OK}}\\
                 $D=1$ & \textcolor{DarkGreen}{\textbf{OK}} & \textcolor{red}{\textbf{error}} & \textcolor{red}{\textbf{error}}\\
             \hline
        \end{tabular}
        \label{tab:tandem-error-cases}
    \end{table}

Therefore, similarly to \eqref{eq:total-error-rate}, the \emph{tandem} total error, $P_\mathscr{E}^\text{tdm}$, is obtained by summing the probabilities of the three (mutually exclusive) errors:
    \begin{equation}
        \begin{aligned}
            P_\mathscr{E}^\text{tdm} = & P(D=0, Y_\text{tdm}=\text{tar}) + \\
            & P(D=1, Y_\text{tdm}=\text{non})+ \\
            & P(D=1, Y_\text{tdm}=\text{spoof})\\
            = & P(D=0|Y_\text{tdm}=\text{tar})P(Y_\text{tdm}=\text{tar})+\\
            & P(D=1|Y_\text{tdm}=\text{non})P(Y_\text{tdm}=\text{non})+\\
            & P(D=1|Y_\text{tdm}=\text{spoof})P(Y_\text{tdm}=\text{spoof})\\
            = & \pi_\text{tar}P_\text{miss}^\text{tdm} + \pi_\text{non}P_\text{fa,non}^\text{tdm}+\pi_\text{spoof}P_\text{fa,spoof}^\text{tdm},
        \end{aligned}\label{eq:total-error-rate-tandem}
    \end{equation}
where
    \begin{equation}
        \begin{aligned}
            P_\text{miss}^\text{tdm}     & \mydef P(D=0|Y_\text{tdm}=\text{tar})\\
            P_\text{fa,non}^\text{tdm}   & \mydef P(D=1|Y_\text{tdm}=\text{non})\\
            P_\text{fa,spf}^\text{tdm} & \mydef P(D=1|Y_\text{tdm}=\text{spoof})\\
        \end{aligned}
    \end{equation}
are, respectively, the miss rate, the false alarm rate for nontargets and the false alarm rate for spoofing attacks. Let us rewrite the total error rate in a more comprehensible form by introducing a new variable 
    \begin{equation}
        \rho \mydef  \frac{\pi_\text{spoof}}{\pi_\text{non}+\pi_\text{spoof}}\nonumber
    \end{equation}
which indicates the proportion of spoofing attacks within the negative class on the evaluation database protocol. We refer to $\rho$ as the \emph{spoof prevalence prior}. We can now rewrite \eqref{eq:total-error-rate-tandem} as, 
    \begin{equation}\label{eq:total-error-rate-convex}
        \begin{aligned}
           P_\mathscr{E}^\text{tdm} & = \pi_\text{tar}P_\text{miss}^\text{tdm} + \pi_\text{non}P_\text{fa,non}^\text{tdm}+\pi_\text{spoof}P_\text{fa,spf}^\text{tdm}\\
            & = \pi_\text{tar}P_\text{miss}^\text{tdm} 
            + \frac{\pi_\text{non}+\pi_\text{spoof}}{\pi_\text{non}+\pi_\text{spoof}}\Bigg[\pi_\text{non}P_\text{fa,non}^\text{tdm}+\pi_\text{spoof}P_\text{fa,spf}^\text{tdm} \Bigg]\\
            & = \pi_\text{tar}P_\text{miss}^\text{tdm} 
            + (\pi_\text{non}+\pi_\text{spoof})\Bigg[(1 - \rho)P_\text{fa,non}^\text{tdm}+
            \rho P_\text{fa,spf}^\text{tdm} \Bigg]\\
            & = \pi_\text{tar}P_\text{miss}^\text{tdm} + (1-\pi_\text{tar})P_\text{fa,$\rho$}^\text{tdm},
        \end{aligned}
    \end{equation}
where the identity $\pi_\text{non}+\pi_\text{spoof}=1-\pi_\text{tar}$ follows from the probabilistic constraints and where 
    \begin{equation}\label{eq:total-tandem-fa-rate}
        P_\text{fa,$\rho$}^\text{tdm} \mydef (1- \rho)P_\text{fa,non}^\text{tdm}+
            \rho P_{\text{fa,spf}}^\text{tdm}
    \end{equation}
denotes the \emph{total tandem false alarm rate}, a convex combination of the two different types of false alarm rates stemming from the acceptance of either nontarget or spoofed trials. For brevity, in the following we refer to \eqref{eq:total-tandem-fa-rate} as the tandem false alarm rate. 

In summary, we re-expressed the total error rate, $P_\mathscr{E}^\text{tdm}$, in terms of the three error rates $(P_\text{miss}^\text{tdm},P_\text{fa,non}^\text{tdm},P_\text{fa,spf}^\text{tdm})$   and the two priors $(\pi_\text{tar},\rho)$. The first prior $\pi_\text{tar}$ expresses the total proportion of the target trials in the database protocol while the latter prior $\rho$ encodes the relative proportion of nontarget and spoofing attack trials corresponding to the negative class. Just as the database prior $\pi$ determines the relative contribution of miss and false alarm rates to the total error rate in the case of a single classifier \eqref{eq:total-error-rate}, \emph{two} database priors are necessary for the tandem scenario to determine the relative contribution of \emph{three} error rates. 

\section{Tandem Equal Error Rate}\label{sec:tandem-equal-error-rate}

We are now equipped with all we need to define formally the tandem equal error rate (t-EER). As in the case of a single classifier, we seek a scalar metric which reflects the overall performance of the tandem system. Further, the metric should be independent not only of decision costs, but also the database prior(s) and should not require the setting of detection thresholds in advance.

\subsection{t-EER}

Just as the classic EER corresponds to an operating point at which the miss and false alarm rates are equal, the \textbf{tandem equal error rate} (t-EER) is characterized by the following equation: 
    \begin{equation}\label{eq:teer-implicit-eq}
        \boxed{P_\text{miss}^\text{tdm} = P_\text{fa,$\rho$}^\text{tdm}}
    \end{equation}
where the miss rate on the left-hand side is again considered a measure of user (in)convenience and the false alarm rate on the right-hand side is a measure of security. The latter combines the two different false alarm rates for nontargets and spoofs. For the time being, the spoof prevalence parameter $\rho$ is considered given (fixed). In Section \ref{sec:concurrent-teer} we describe an approach to estimate a tandem error rate that is, like the conventional EER, independent of the class priors.

It it important to note that while the classic EER is \emph{always} a unique scalar that corresponds \emph{typically} to a unique EER threshold, neither holds for the t-EER. In practice, there will be multiple operating points which satisfy \eqref{eq:teer-implicit-eq}, each with different values of $P^\text{tdm}_\text{miss}=P^\text{tdm}_\text{fa,$\rho$}$, leading to a \textbf{t-EER path} $\Omega(\rho)$ for any given $0 \leq \rho \leq 1$. Formally,
    \begin{equation}\label{eq:tEER-path}
        \boxed{
        \Omega(\rho) \mydef \{\vec{\tau}\in\mathbb{R}^2 : P_\text{miss}^\text{tdm}(\vec{\tau}) = P_\text{fa,$\rho$}^\text{tdm}(\vec{\tau})\}}
    \end{equation}
where $\vec{\tau}=(\tau_\text{asv},\tau_\text{cm})$ denote now a \emph{pair} of subsystem thresholds. While further details will be provided below, the eager reader may already take a sneak peek of Figs. \ref{fig:critical-thresholds}, \ref{fig:tEER_example_2d}, and \ref{fig:tEER_example_2d_real_scores} to see how typical t-EER paths looks like. 

\subsection{Decision Regions for the AND rule}

Thus far we have not assumed any particular form in which CM and ASV subsystems are combined. Let us now invoke the assumption in \eqref{eq:AND-combination}. By letting $D(\vec{s})$ denote the tandem decision for a particular input score pair $\vec{s}=(s_\text{asv},s_\text{cm})$, the `accept' and `reject' decision regions are given by
    \begin{equation}\label{eq:accept-reject-sets}
        \begin{aligned}
            \mathsf{ACCEPT} & = \{\vec{s} \in \mathbb{R}^2: D(\vec{s})=1\} = A_1\\
            \mathsf{REJECT} & = \{\vec{s} \in \mathbb{R}^2: D(\vec{s})=0\} = A_2 \cup A_3 \cup A_4,\\
        \end{aligned}
    \end{equation}
as illustrated in Fig. \ref{subfig:areas}. This classification rule is defined by two orthogonal hyperplanes which partition $\mathbb{R}^2$ into four non-overlapping connected regions labeled $A_1$ through $A_4$. A test trial, represented by a pair of scores $\vec{s}$, is accepted if (and only if) $\vec{s} \in A_1$. Score vectors $\vec{s}$ that fall in either one of the other three regions are rejected. 

\subsection{Tandem Miss and False Alarm rates Under the AND Rule}\label{subsec:tandem-miss-and-FA-AND-rule}

To proceed algebraically from \eqref{eq:teer-implicit-eq}, let us begin from the left-hand side. A miss takes place whenever a positive trial (bona fide target) represented by a pair of scores $\vec{s}=(s_\text{asv},s_\text{cm}) \in \mathbb{R}^2$ falls into the union of $A_2$, $A_3$ and $A_4$. This region could be formed in different ways, such as $(A_2 \cup A_4) \cup A_3$ or $(A_3 \cup A_4) \cup A_2$. Following the former, the miss rate is
    \begin{equation}
        \begin{aligned}\nonumber
            P_\text{miss}^\text{tdm} & = \underbrace{P(S_\text{asv} \leq \tau_\text{asv}|\text{tar})}_{P(\text{target}\; \in A_2 \cup A_4)} + \underbrace{P(S_\text{asv} > \tau_\text{asv}|\text{tar})P(S_\text{cm} \leq \tau_\text{cm}|\text{tar})}_{P(\text{target}\; \in A_3)}\\
            & = P_\text{miss}^\text{asv}(\tau_\text{asv})+\big(1-P_\text{miss}^\text{asv}(\tau_\text{asv})\big)P_\text{miss}^\text{cm}(\tau_\text{cm}),\\
        \end{aligned}
    \end{equation}
where $P_\text{miss}^\bullet(\tau_\bullet)$ denotes the subsystem-specific miss rates at their respective thresholds. The miss rate consists of two terms, the union of which gives rise to the yellow shaded area in Fig~\ref{subfig:missfa}. The first term is given by equation (ii) illustrated to the bottom of Fig~\ref{subfig:missfa}. (ii), while the second term is the product of (i) and (iv). 

\emph{Alternatively}, one could also obtain the same shaded area by combining (iv) with the product of (ii) and (iii), as follows
    \begin{equation}
        \begin{aligned}\nonumber
            P_\text{miss}^\text{tdm} & = \underbrace{P(S_\text{cm} \leq \tau_\text{cm}|\text{tar})}_{P(\text{target}\; \in A_3 \cup A_4)} + \underbrace{P(S_\text{asv} \leq \tau_\text{asv}|\text{tar})P(S_\text{cm} > \tau_\text{cm}|\text{tar})}_{P(\text{target}\; \in A_2)}\\ 
            & = P_\text{miss}^\text{cm}(\tau_\text{cm}) + P_\text{miss}^\text{asv}(\tau_\text{asv}) \big(1 - P_\text{miss}^\text{cm}(\tau_\text{cm})\big)
        \end{aligned}
    \end{equation}
It is clear that either approach yields the same result:
    \begin{equation}\label{eq:tandem-miss-rate}
        \begin{aligned}
            P_\text{miss}^\text{tdm} 
            & = P_\text{miss}^\text{cm}(\tau_\text{cm})+P_\text{miss}^\text{asv}(\tau_\text{asv})-P_\text{miss}^\text{cm}(\tau_\text{cm})P_\text{miss}^\text{asv}(\tau_\text{asv}).
        \end{aligned}
    \end{equation}
Having determined the expression for tandem miss rate, we proceed with the right-hand side of \eqref{eq:teer-implicit-eq}. It consists of the convex sum of the two different types of false alarms in \eqref{eq:total-tandem-fa-rate}. The two different errors take place, respectively, when a nontarget trial or a spoof trial falls into $A_1$. For the former, we have
\begin{equation}
\begin{aligned}
P_{\text{fa,non}}^{\text{tdm}} & = \underbrace{P(S_{\text{cm}} > \tau_{\text{cm}}|\text{bona})P(S_{\text{asv}} > \tau_{\text{asv}}|\text{non})}_{P(\text{nontarget}\; \in A_1)}\\
& = \big(1-F_{S_{\text{cm}}}(\tau_{\text{cm}}|\text{bona})\big)\big(1-F_{S_{\text{asv}}}(\tau_{\text{asv}}|\text{non})\big)\\
& = \big(1 - P_{\text{miss}}^{\text{cm}}(\tau_{\text{cm}})\big)P_{\text{fa,non}}^{\text{asv}}(\tau_{\text{asv}}).
\end{aligned}
\end{equation}

Here, we assert $p(S_\text{cm}|\text{bona}):= p(S_\text{cm}|\text{tar})=p(S_\text{cm}|\text{non})$. Since both target and nontarget correspond to bona fide (human) data, their corresponding CM score distribution are similar. We point the interested reader to \cite[Section III.B]{Kinnunen2020-tandem-fundamentals} for further discussion. Finally, for spoofing attacks
    \begin{equation}
        \begin{aligned}\nonumber
            P_\text{fa,spoof}^\text{tdm} & = \underbrace{P(S_\text{cm} > \tau_\text{cm}|\text{spoof})P(S_\text{asv} > \tau_\text{asv}|\text{spoof})}_{P(\text{spoofing attack}\; \in A_1)}\\
& = \big(1-F_{\text{S}_{\text{cm}}}(\tau_{\text{cm}}|\text{spoof})\big)\big(1-F_{\text{S}_{\text{asv}}}(\tau_{\text{asv}}|\text{spoof})\big)\\
            & = P_\text{fa}^\text{cm}(\tau_\text{cm})P_\text{fa,spoof}^\text{asv}(\tau_\text{asv})
        \end{aligned}
    \end{equation}

\label{correlation}
Note that in formulating the tandem miss and false acceptance rates above, we have explicitly assumed that the ASV and CM scores are \emph{conditionally independent} such that $p(s_\text{asv}, s_\text{cm} | Y_\text{tdm}) = p(s_\text{asv}|Y_\text{tdm})p(s_\text{cm}|Y_\text{tdm})$.
In biometric verification this is a reasonable assumption which greatly simplifies the computation of the tandem error rates (i.e., the shaded area in Fig.~\ref{subfig:missfa}) as the product of the error rates evaluated on the ASV and CM score distributions, respectively. While an ASV system focuses on the detection of voice attributes, the CM system aims to discover artifacts (see Appendix for empirical analysis on correlation of several state-of-the-art ASV and CM systems), thereby justifying the independence assumption; see also~\cite[Section VIII.A]{Kinnunen2020-tandem-fundamentals}. Note \emph{class-conditional} in the independence assumption. Without conditioning, the ASV and CM scores tend to exhibit stronger correlation, as seen from typical empirical scores in Fig.~\ref{fig:tEER_example_2d_real_scores} as well as from the sketch in Fig.~\ref{subfig:missfa}. However, this global ASV-CM score dependency is suppressed once conditioned on the class.

All three tandem error rates are functions of the two thresholds, $\vec{\tau}=(\tau_\text{asv},\tau_\text{cm})$. Let us first examine the error rates at the limits where $\vec{\tau} \in \{(\pm \infty, \pm \infty) \}$, as listed in Table \ref{tab:edge-cases}. These values follow directly from the properties of the CDF. The last column also shows the tandem false alarm rate defined in \eqref{eq:total-tandem-fa-rate}. Note that, for these limit cases, its value is independent of $\rho$.

\begin{table}[]
    \centering
    \caption{Values of tandem miss rate, tandem false alarm, tandem spoof false alarm and total false alarm rates at the limit cases.}
    \begin{tabular}{c|cccc}
         $\vec{\tau}=(\tau_\text{asv},\tau_\text{cm})$ & $P_\text{miss}^\text{tdm}$ & $P_\text{fa,non}^\text{tdm}$ & $P_\text{fa,spf}^\text{tdm}$ & $P_{\text{fa},\rho}^\text{tdm}$ \\ \hline\hline
         $(-\infty,-\infty)$ &  0 & 1 & 1 & 1\\
         $(-\infty,+\infty)$ &  1 & 0 & 0 & 0\\
         $(+\infty,-\infty)$ &  1 & 0 & 0 & 0\\
         $(+\infty,+\infty)$ &  1 & 0 & 0 & 0\\
         \hline
    \end{tabular}
     \label{tab:edge-cases}
\end{table}

As for the conventional EER, the total miss and false alarm rates have contrasting 0/1 values for each pair of threshold limits in the first column of Table \ref{tab:edge-cases}. For instance, a comparison of the first and third rows reveals that, when the CM threshold is set to $-\infty$, the tandem miss rate is 1 for an ASV threshold of $+\infty$ and 0 for an ASV threshold of $-\infty$. The first row corresponds to an accept-all condition, while the remaining rows correspond to a reject-all condition. Whenever either threshold is set to $+\infty$, either $S_\text{cm} \leq +\infty$ and/or $S_\text{asv} \leq +\infty$ hold. The \texttt{AND} rule then always leads to rejection (by the tandem system), giving miss rate of 1 and false alarm rate(s) of 0.

\begin{table}[!t]
    \centering
    \caption{A unified summary of the commonly-used EERs in speaker verification and spoofing detection (first column) along with the respective positive and negative class definitions, along with the ASV ($\tau_\text{asv}$) and countermeasure ($\tau_\text{cm}$) thresholds and spoof prevalence ($\rho$) of the evaluation database. Here `$*$' indicates the free threshold(s) that are varied to find the operating point(s) at which miss rate equals false alarm rate.}
    \begin{tabular}{c|cc|c}
         The type of EER & Positive & Negative & $(\tau_\text{asv},\tau_\text{cm},\rho)$\\
         \hline\hline
         ASV (tar vs non) & target & nontar.                 & $(*, -\infty, 0)$\\
         ASV (tar vs spf) & target & spoof  & $(*, -\infty, 1)$\\
         ASV (tar vs non+spf) & target & nontar. $\cup$ spoof & $(*, -\infty, \rho)$\\
         CM (bona vs spf) & bona fide & spoof & $(-\infty, *, 1)$\\
         \textbf{t-EER of (ASV,CM)} & target & nontar. $\cup$ spoof & $(*,*,\rho)$\\
         \hline
    \end{tabular}
    \label{tab:summary-of-tEER-special-cases}
\end{table}

\subsection{t-EER of Accept-All and Reject-All Subsystems}\label{subsec:tEER-acceptall-rejectall}

With the above definitions in place, let us first consider a few special systems and database protocols. First, note that the t-EER is \emph{undefined} for all the four cases listed in the rows of Table \ref{tab:edge-cases}; since the tandem miss rate and the tandem false alarm rate have opposing values, they cannot be made equal.

\textbf{Accept-all ASV.} Now consider the case when the ASV sub-system accepts everything ($\tau_\text{asv}=-\infty)$, while $|\tau_\text{cm}| < \infty$. In this case the tandem system reduces to a CM-only system handling three classes of trials:
    \begin{equation}
        \begin{aligned}
            P_\text{miss}^\text{tdm} & = P_\text{miss}^\text{cm}(\tau_\text{cm})\nonumber\\
            P_\text{fa,$\rho$}^\text{tdm} & = (1-\rho)(1-P_\text{miss}^\text{cm}(\tau_\text{cm}))+\rho P_\text{fa}^\text{cm}(\tau_\text{cm})  
        \end{aligned}
    \end{equation}
For the special case $\rho=0$ (i.e.\ there are no spoofed trials), the tandem false alarm rate is $P_\text{fa,$\rho$}^\text{tdm}=1-P_\text{miss}^\text{cm}(\tau_\text{cm})$. As per \eqref{eq:teer-implicit-eq}, we determine the t-EER by equating the two tandem error rates which gives $P_\text{miss}^\text{cm}(\tau_\text{cm})=1-P_\text{miss}^\text{cm}(\tau_\text{cm})$, i.e. $P_\text{miss}^\text{cm}(\tau_\text{cm})=\frac{1}{2}$. That is, 
the CM threshold should be set so that the CM miss rate is $\frac{1}{2}$, or 50\%. This is the case also for the CM false alarm rate, giving a t-EER of $\frac{1}{2}$. For $\rho=1$ (i.e.\ there are now no nontarget trials), the t-EER is given by $P_\text{miss}^\text{cm}(\tau_\text{cm})=P_\text{fa}^\text{cm}(\tau_\text{cm})$. The CM threshold $\tau_\text{cm}$ must now be set so that the CM miss and false alarm rates are equal --- hence, the t-EER is the familiar CM EER.

\textbf{Accept-all CM.} We now consider a similar treatment, but for an accept-all CM (i.e.\ $\tau_\text{cm}=\textcolor{blue}{-}\infty$), while $|\tau_\text{asv}| < \infty$. Now the tandem system reduces to a ASV-only system:
    \begin{equation}
        \begin{aligned}
            P_\text{miss}^\text{tdm} & = P_\text{miss}^\text{asv}(\tau_\text{asv})\nonumber\\
            P_\text{fa,$\rho$}^\text{tdm} & = (1-\rho)P_\text{fa,non}^\text{asv}(\tau_\text{asv})+\rho P_\text{fa,spf}^\text{asv}(\tau_\text{asv}).  
        \end{aligned}
    \end{equation}

Let us find again the t-EER by equating the tandem miss and false alarm rates. For the special case $\rho=0$, the ASV threshold $\tau_\text{asv}$ should be set so that $P_\text{miss}^\text{asv}(\tau_\text{asv})=P_\text{fa,non}^\text{asv}(\tau_\text{asv})$. This is the conventional ASV EER. Similarly, $\rho=1$ gives the EER of the ASV system, but with no nontarget trials. For the intermediate values $0 < \rho < 1$, the t-EER becomes the EER of an ASV system computed using a mix of nontarget and spoof trials. This EER was referred to as the \emph{spoof-aware speaker verification} (SASV) EER in~\cite{SASV2022-eval-plan}.\footnote{To be specific, the particular value of $\rho$ (computed from the evaluation protocol) for the SASV challenge is $\rho \approx 0.65$ --- even if not reported in this form.} Thus, \emph{this metric depends on the empirical composition of spoofed and non-target trials (i.e., the spoof prevalence prior) in the test set}, making the SASV EER ill-suited for performance comparisons made across different protocols or databases with a different empirical composition.

A summary of the special case t-EERs is shown in Table \ref{tab:summary-of-tEER-special-cases}. The four familiar EERs used in speaker verification and spoofing detection --- the EER of the CM and the three different EERs of an ASV system (with nontargets, spoofs and a mix of both) are all special cases of the t-EER with specific subsystems and database assumptions. \textbf{Hence, the t-EER is a unifying concept: whenever one computes and reports one of the familiar per-subsystem EERs, it is equivalent to reporting t-EERs with special database protocol or subsystem assumptions}.

\begin{figure}
    \centering
    \includegraphics[width=0.98\linewidth]{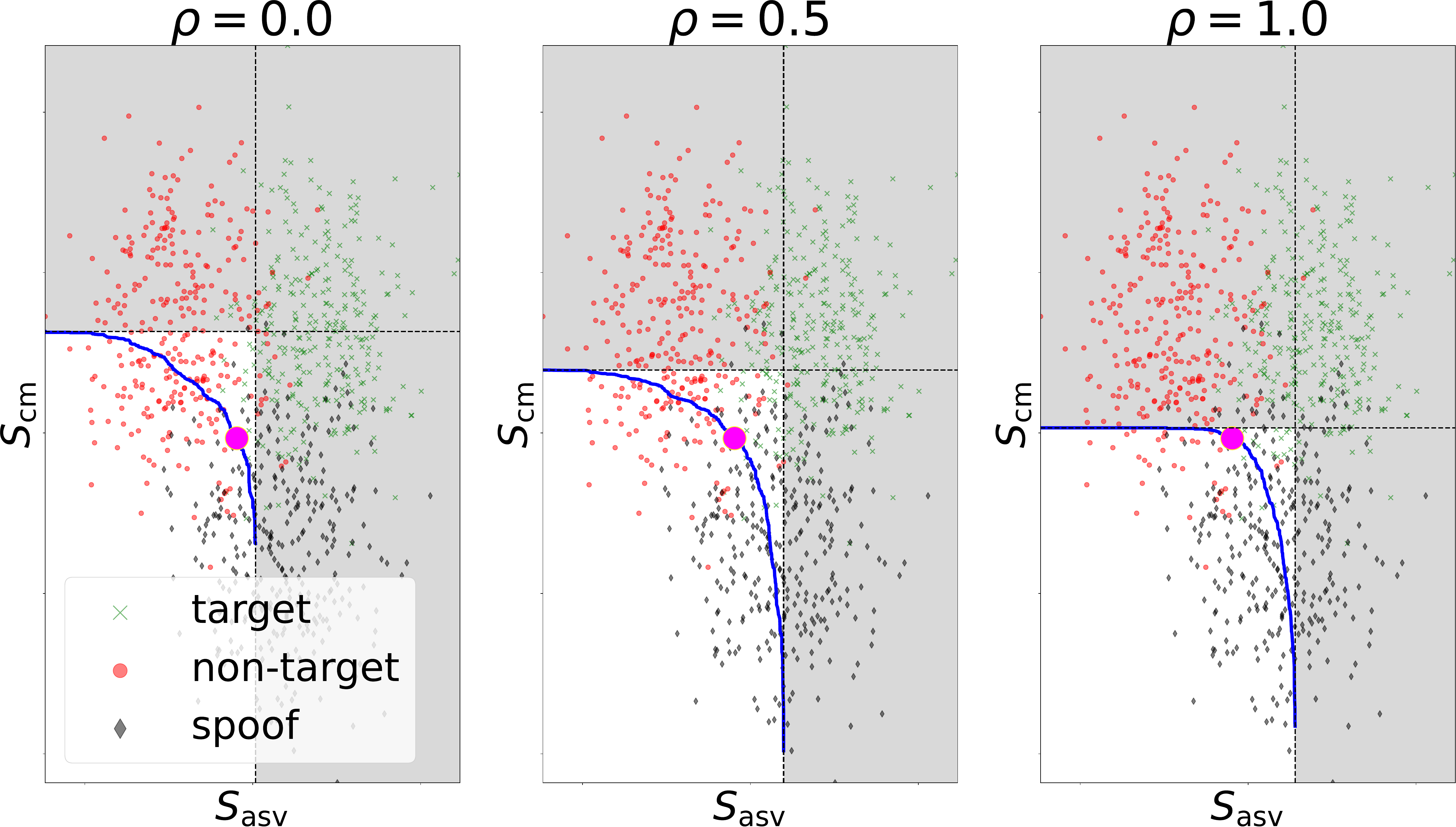}
    \caption{The t-EER path (blue solid lines) for a given spoofing prevalence prior $\rho$ lives in the constrained region of $\mathbb{R}^2$ (unshaded) defined by two critical thresholds $\tau^\text{asv}_*$ and $\tau^\text{cm}_*$ (dashed black lines). The expressions for the respective critical thresholds are provided by Eq. \eqref{eq:asv-feasible-region} (ASV) and \eqref{eq:cm-feasible-region} (CM).}
    \label{fig:critical-thresholds}
\end{figure}

\section{Computation of the t-EER Path}\label{sec:teer-computation}

We now turn our attention toward the case of non-degenerate subsystems with $-\infty < \tau_\text{asv},\tau_\text{cm} < +\infty$ where the practical computation of the t-EER path requires more effort. To this end, the t-EER path can be found by considering one of the thresholds fixed (in an outer loop) while sweeping through values of the other (inner loop) in order to determine threshold pairs for which the tandem miss and false alarm rates have the same (or similar\footnote{Due to their discrete nature, the tandem miss and false alarm rates may not be precisely identical. Nonetheless, since one is monotonically increasing and the other monotonically decreasing, their profiles will intersect at some operating points defined by a pair of thresholds $\tau_\text{cm}$ and $\tau_\text{asv}$.}) values. Nonetheless, the fixed threshold cannot be set arbitrarily. In the following we provide the technical conditions alongside the visual aid in Fig. \ref{fig:critical-thresholds}.

\subsection{The t-EER path lives in the lower-left score region}

We begin by considering (arbitrarily) a fixed ASV threshold and a variable CM threshold. The following result states that the ASV threshold cannot be selected arbitrarily.
\begin{lemma}
    \label{lemma:critical-asv}
    Let $0 \leq \rho \leq 1$ be arbitrary and let $\tau^\text{asv}_* \in \mathbb{R}$ be selected so that
        \begin{equation}\label{eq:asv-feasible-region}
            (1 - \rho)P_\text{fa,non}^\text{asv}(\tau^\text{asv}_*) + \rho P_\text{fa,spoof}^\text{asv}(\tau^\text{asv}_*) \geq P_\text{miss}^\text{asv}(\tau^\text{asv}_*)
        \end{equation}
    holds (i.e., the total false alarm rate of ASV at $\tau^\text{asv}_*$ is at least as high as the ASV miss rate). Then there exists $\tau_\text{cm} \in \mathbb{R}$ so that $(\tau^\text{asv}_*,\tau_\text{cm}) \in \Omega(\rho)$. Likewise, if \eqref{eq:asv-feasible-region} does \emph{not} hold, then there is \emph{no} such $\tau_\text{cm}$ for which $(\tau^\text{asv}_*,\tau_\text{cm}) \in \Omega(\rho)$.
\end{lemma}

The proof is provided in the Appendix. As illustrated in Fig.~\ref{fig:critical-thresholds}, the essence of \eqref{eq:asv-feasible-region} is that the t-EER exists only when the ASV threshold is \emph{sufficiently low}. The CM cannot then be configured to lower the \emph{tandem} miss rate to the same value as the false alarm rate, even by setting $\tau_\text{cm}=-\infty$. In particular, note that if $\tau^\text{asv}_* = +\infty$ (i.e.\ `a reject-all' ASV system), the requirement \eqref{eq:asv-feasible-region} is
    \begin{equation}
        (1-\rho)\cdot 0 + \rho \cdot 0 \geq 1 \Leftrightarrow 0 \geq 1,
    \end{equation}
which is identically false. In contrast, for $\tau_*^\text{asv} = -\infty$ the condition becomes identically true ($0 \leq 1$). The range of ASV thresholds for which the t-EER exists are those lower than the critical threshold that makes the left and right hand sides of \eqref{eq:asv-feasible-region} equal.

\begin{figure}
    \centering
     \includegraphics[trim={2.3cm 2.5cm 3cm 2cm},clip,width=1.15\linewidth]{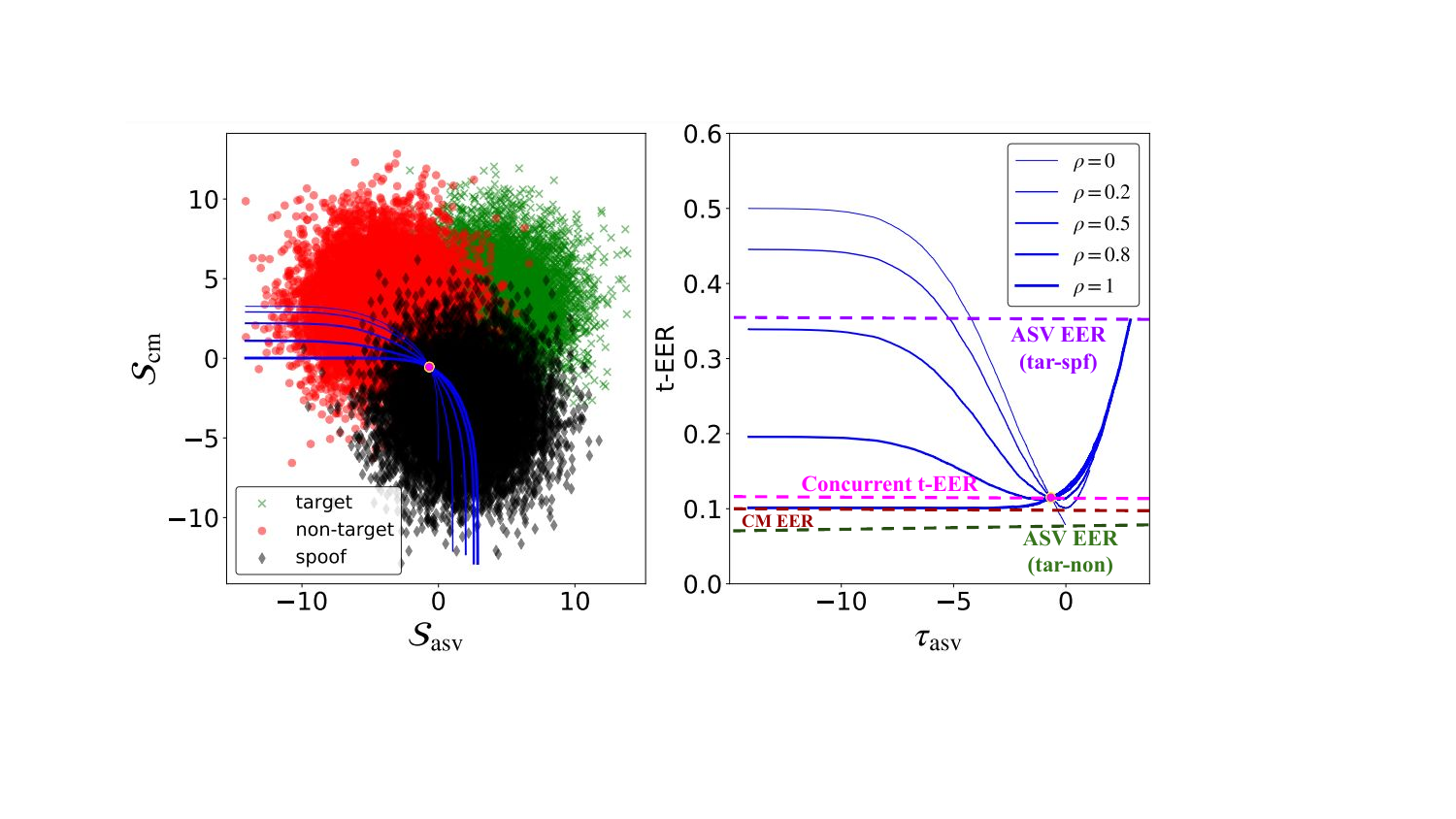}
    \caption{Demonstration of t-EER paths and t-EER values for different values of $\rho$ on simulated scores drawn from three bivariate Gaussians. The overlaid blue curves on the left-hand side display the t-EER paths (one for each $\rho$), while the corresponding curves on the right-hand side display the corresponding t-EER \emph{value} along each path. Three familiar special-case EERs are also indicated, along with proposed \emph{concurrent} t-EER (defined in this work).}
    \label{fig:tEER_example_2d}
\end{figure}

\begin{figure}
    \centering
    \includegraphics[trim={4.8cm 2.5cm 2.5cm 2.5cm},clip,width=1.03\linewidth]{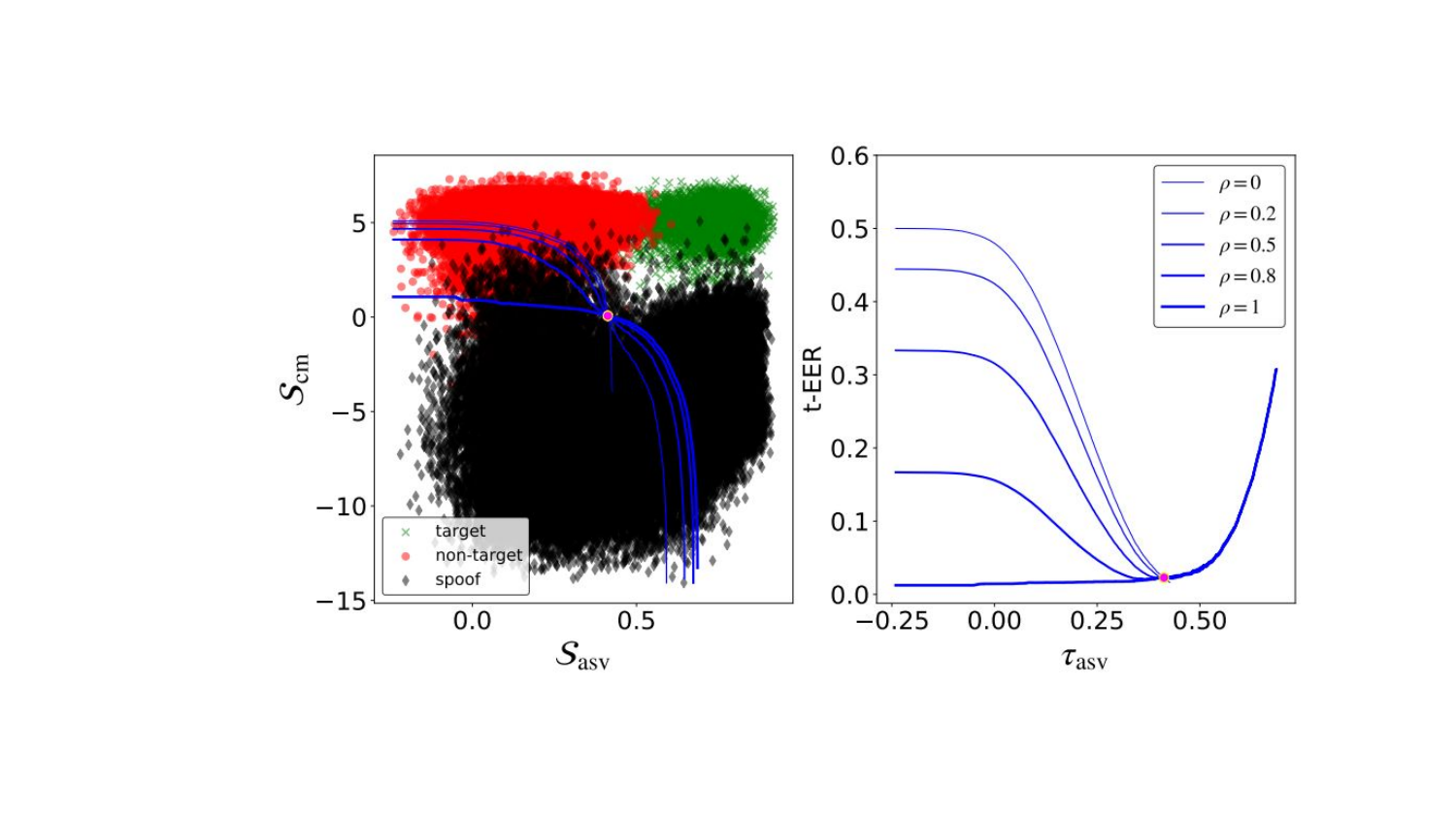}
    \caption{Demonstration of t-EER paths and t-EER values for different values of $\rho$ on empirical scores from ASV and CM systems (SASV baseline B1). The overlaid blue curves on the left-hand side display the t-EER paths (one for each $\rho$), while the corresponding curves on the right-hand side display the corresponding t-EER \emph{value} along each path. }
    \label{fig:tEER_example_2d_real_scores}
\end{figure}
The following dual lemma concerns the existence of the t-EER when the CM threshold is fixed, instead of the ASV threshold.

\begin{lemma}
    Let $0 \leq \rho \leq 1$ be arbitrary and let $\tau^\text{cm}_* \in \mathbb{R}$ be the CM threshold selected so that
        \begin{equation}\label{eq:cm-feasible-region}
            1-\rho + \rho P_\text{fa}^\text{cm}(\tau^\text{cm}_*) \geq (2 - \rho)P_\text{miss}^\text{cm}(\tau^\text{cm}_*) 
        \end{equation}
holds. Then there exists $\tau_\text{asv} \in \mathbb{R}$ so that $(\tau_\text{asv},\tau^\text{cm}_*) \in \Omega(\rho)$.
\end{lemma}
\noindent Proof is similar to that of Lemma \ref{lemma:critical-asv} (omitted). $\blacksquare$

The result is similar. For the t-EER to exist, the CM threshold should be set low enough in order to ensure a sufficiently high false alarm rate. This allows for the ASV threshold to be set high enough so as to achieve an equal tandem miss rate. Taken together, Eq. \eqref{eq:asv-feasible-region} and \eqref{eq:cm-feasible-region} provide useful empirical maximum values for the ASV and CM thresholds that need to be considered in the computation of the t-EER path.

\subsection{Construction of the t-EER Path}

Up to this point, we have purposefully distanced ourselves from practicalities of the t-EER path construction. Nonetheless, this is an important consideration, particularly in terms of memory consumption. In a practical situation, we have a pair of detection scores, each one generated by one of the two subsystems.
    \begin{equation}
        \begin{aligned}
            \mathcal{S}_\text{asv} & = \Big(s_\text{asv}^{(1)},\dots,s_\text{asv}^{(N)}\Big)\nonumber\\
            \mathcal{S}_\text{cm} & = \Big(s_\text{cm}^{(1)},\dots,s_\text{cm}^{(M)}\Big),
        \end{aligned}
    \end{equation}
along with their respective ground-truth labels $y_\text{asv}^{(i)}, i=1,\dots,N$ and $y_\text{cm}^{(j)} ,j=1,\dots,M$. Each score set individually takes up linear space in the number of trials and is neglible on modern computer systems. Even if usually not the case, the ASV and CM scores may originate from different sets of test cases, with $N \neq M$. Using the two sets of detection scores one may then construct empirical class-conditional CDFs per subsystem using sorting procedures \cite{Fawcett2005-an-intro-to-ROC-analysis}. As known from conventional single-system ROC analysis (e.g. \cite{ROC-curves-2009}), the number of distinct pairs of miss and false alarm rate values is the number of unique score values plus one. Hence, the resulting empirical tandem miss and false alarm rates (which consist of all pairs of the two thresholds) can be presented as a 2D array of shape $(N+1) \times (M+1)$ which takes quadratic memory for the usual case $N=M$. As typical biometrics and spoofing detection evaluation benchmarks consist of hundreds of thousands to tens of millions of evaluation trials, it is clear that storing the full 2D table becomes prohibitively impractical even for modern computer systems.

Since we are interested only in the t-EER \emph{path}, we do not need to store all these values but, rather, a 1D array of table indices $\texttt{idx}[i], 1 \leq i \leq N+1$ where $1 \leq \texttt{idx}[i] \leq M+1$ contains the index of CM threshold corresponding to an ASV threshold indexed by $i$. Here, each $[i, \texttt{idx}[i]]$ indices a point on the (now discretely-defined) t-EER path. Depending on whether there are more ASV or CM trials, one may decide to store either the CM threshold indices (one per each ASV threshold that fulfills \eqref{eq:asv-feasible-region}); or the ASV threshold indices (one per each CM threshold that fulfills \eqref{eq:cm-feasible-region}). Either case leads to linear, rather than quadratic, memory usage.

\section{The concurrent t-EER:\\ a Unique Scalar Summary}\label{sec:concurrent-teer}

As noted above, the conventional EER is uniquely defined for a given set of scores. Unfortunately, this is generally \emph{not} the case for the t-EER defined in \eqref{eq:teer-implicit-eq}. That is, if $\vec{\tau}_1,\vec{\tau}_2 \in \Omega(\rho)$ are two distinct points ($\vec{\tau}_1 \neq \vec{\tau}_2$) on a t-EER path, then the t-EERs at $\vec{\tau}_1$ and $\vec{\tau}_2$ correspond to \emph{different} values of $\text{t-EER}=P_\text{miss}^\text{tdm}=P_\text{fa,$\rho$}^\text{tdm}$. Hence, the t-EER is in fact not a scalar but a \emph{function} with the t-EER path $ \Omega(\rho)$ as its domain. 

A natural question arises: \emph{which} t-EER value along the t-EER path should we choose as a meaningful summary statistic? Given the optimum (oracle) threshold sentiment of conventional EER, it may seem very reasonable to consider the \emph{minimum}. Unfortunately, since the t-EER path (and the corresponding t-EER values) depends on $\rho$, that minimum will also depend on $\rho$. This leads to the same empirical trial composition dependency problem noted in Subsection \ref{subsec:tEER-acceptall-rejectall}. Our answer to the original question will be \emph{choose the unique t-EER value shared across \emph{all} spoof prevalence priors, at the unique intersection point of all possible paths $\{\Omega(\rho): \rho \in [0,1]\}$.} 

Before formalizing our proposal, let us consider a simple example. Fig.~\ref{fig:tEER_example_2d} shows an illustrative example of t-EER analysis for a set of simulated CM and ASV scores sampled from three bivariate normal distributions, one per each of the three classes (target, non-target and spoof).\footnote{Simulated scores are produced similarly to \cite[Appendix]{Kinnunen2020-tandem-fundamentals} where the marginals define the ASV and CM score distributions parameterized through their respective EERs. In this simulation, we use arbitrary EER values of 8\% (ASV target-vs-nontarget); 35\% (ASV target-vs-spoof); and 10\% (CM target-vs-spoof), and diagonal covariance matrices.} The t-EER paths are derived from the treatment presented above and are reproducible using online code\footnote{\label{fn:url}\url{https://colab.research.google.com/drive/1ga7eiKFP11wOFMuZjThLJlkBcwEG6_4m?usp=sharing}} 
Profiles are plotted for three different values of the spoofing prevalence prior $\rho$. The plot to the left shows CM scores (vertical axis) and ASV scores (horizontal axis) for target trials (green), non-target trials (red) and spoofed trials (black).  Also shown are superimposed t-EER paths for each $\rho$ defined by corresponding $\tau_\text{cm}$ and $\tau_\text{asv}$ threshold pairs. The plot to the right in Fig. \ref{fig:tEER_example_2d} shows the t-EER as a function of the ASV threshold $\tau_\text{asv}$.

Recall that the spoof prevalence parameter $\rho$ combines the two tandem false alarm rates in \eqref{eq:total-tandem-fa-rate} corresponding to the non-target and spoof classes, respectively. The t-EER paths equate the tandem miss rate to the non-target and spoof tandem false alarm rates at different thresholds. The value of $\rho$ determines the prevalence of the spoofed class, with respect to the non-target class, in the tandem false rejection rate. For lower $\rho$, the nontarget class has a larger weight than the spoofed class. As such, the t-EER is higher for lower $\tau_\text{asv}$ (higher non-target false acceptance rate) but lower for higher $\tau_\text{asv}$ as noted in Fig.~\ref{fig:tEER_example_2d}.

As expected, the profiles show that the value of the t-EER differs along each path for all three values of the spoofing prevalence prior $\rho$.  Of specific interest is the apparent intersection of each path at what appears to be the same values of $\tau_\text{cm}$ and $\tau_\text{asv}$.  At this specific operating point, the value of the t-EER also appears to be identical and is, hence, independent of $\rho$.

\subsection{The concurrent t-EER: where t-EER paths intersect}

Up until now, our treatment has concerned t-EER \emph{paths} $\Omega(\rho)$, rather than a single scalar like the conventional EER. Each path characterises a set of operating points that satisfy \eqref{eq:teer-implicit-eq} and for which the t-EER may be different. In further contrast to the conventional EER, the paths are dependent on the database prior, $\rho$.  This dependence is undesirable, not least because it implies that estimates of performance derived from different databases cannot be made using the t-EER; the estimates are dependent on the database or, more specifically, the spoofing prevalence prior for that database. 

As shown in Fig.~\ref{fig:tEER_example_2d}, the dependence on the database prior is seen for each point along each t-EER path, except for the particular operating point at which the t-EER paths intersect. At this particular point, the t-EER is independent of $\rho$ just like the conventional EER.

\begin{theorem} (Characterization of t-EER path intersection) \label{thm:concurrent-eer}
Let $0 \leq \rho_1, \rho_2 \leq 1$ be arbitrary but distinct ($\rho_1 \neq \rho_2$) spoof prevalence priors. Then the corresponding t-EER paths $\Omega(\rho_1)$ and $\Omega(\rho_2)$ intersect at $\vec{\tau}^\times \mydef (\tau_\text{asv}^\times,\tau_\text{cm}^\times)$, where the two thresholds satisfy
    \begin{equation}\label{eq:xpoint-equation}
        \frac{P_\text{fa,non}^\text{asv}(\tau_\text{asv}^\times)}{P_\text{fa,spf}^\text{asv}(\tau_\text{asv}^\times)} = \frac{P_\text{fa}^\text{cm}(\tau_\text{cm}^\times)}{1 - P_\text{miss}^\text{cm}(\tau_\text{cm}^\times)}.
    \end{equation}
Further, the t-EER at $\vec{\tau}_\times$, denoted by $P_\mathscr{E}^\times$, is
    \begin{equation}\label{eq:t-EER-at-xpoint}
        P_\mathscr{E}^\times = P_\text{fa,spf}^\text{asv}(\tau_\text{asv}^\times)P_\text{fa}^\text{cm}(\tau_\text{cm}^\times). 
    \end{equation}
\end{theorem}

\noindent The proof (see Appendix A) shows that the t-EER paths for different values of $\rho$ have a common intersection point and a common value of t-EER on that point. For simplicity, consider the boundary cases for the spoofing prevalence prior $\rho=0$ (no spoofing attacks present) and $\rho=1$ (no non-targets present). Their t-EER at the intersection point is given by
\begin{equation}
    \begin{aligned}
            P_\text{miss}^\text{tdm}(\vec{\tau}^\times) & = P_\text{fa,non}^\text{tdm}(\vec{\tau}^\times) & {\quad\text{for\quad} \rho = 0}\\
            P_\text{miss}^\text{tdm}(\vec{\tau}^\times) & = P_\text{fa,spoof}^\text{tdm}(\vec{\tau}^\times) & {\quad\text{for}\quad \rho = 1}
    \end{aligned}
    \label{eq:concurent_t-eer_boundary}
\end{equation}
since $ P_\text{miss}^\text{tdm}(\vec{\tau}^\times)$ is, by definition, the same for all $\rho\in[0,1]$. From the above equations, it can be seen that, at the point of intersection, the three error rates are \emph{concurrently} equal:
\begin{equation}
    \boxed{
    P_\mathscr{E}^\times = P_\text{miss}^\text{tdm}(\vec{\tau}_\times) = P_\text{fa,non}^\text{tdm}(\vec{\tau}^\times) = P_\text{fa,spoof}^\text{tdm}(\vec{\tau}^\times)}
    \label{eq:concurent_t-eer}
\end{equation}
We refer to the value of the t-EER at the point of intersection as the \textbf{concurrent t-EER}, which uniquely characterizes the performance of a tandem system with a single parameter-free performance metric. It is worth noting that one could also deduce \eqref{eq:concurent_t-eer} from \eqref{eq:crossover}[Appendix A] using the expression in the parentheses. It is worth noticing that the concurrent relation in \eqref{eq:concurent_t-eer} indicates a unique point $ P_\mathscr{E}^\times$ on the CM-ASV score space where the target, non-target, and spoofed score distributions intersect with the three error rates being equal. The boundary conditions (where either nontarget or spoof class vanishes) are given by \eqref{eq:concurent_t-eer_boundary}.

\subsection{The relationship between the  concurrent t-EER, the total error rate and the t-DCF} 

Let us denote the concurrent t-EER value by $P_\mathscr{E}^\times$. Straightforward substitution of $P_\mathscr{E}^\times$ in place of $P_\text{miss}^\text{tdm}$, $P_\text{fa,non}^\text{tdm}$, and $P_\text{fa,$\rho$}^\text{tdm}$ in \eqref{eq:total-error-rate-tandem} shows that $P_\mathscr{E}^\text{tdm}=P_\mathscr{E}^\times$, i.e. the concurrent t-EER and total tandem error rate equal one another (at $\vec{\tau}_\times$). This is a direct extension of the relationship between the classical total error rate and the EER discussed in Section~\ref{subsec:classic-EER}.

Another relevant connection can be made to the tandem detection cost function (t-DCF) \cite{Kinnunen2020-tandem-fundamentals}, which has the form
    \begin{equation}\label{eq:tdcf-eq}
        \text{t-DCF} = C_\text{miss}\pi_\text{tar}'P_\text{miss}^\text{tdm}+C_\text{fa,non}\pi_\text{non}'P_\text{fa,non}^\text{tdm}+C_\text{fa,spf}\pi_\text{spf}'P_\text{fa,spf}^\text{tdm},
    \end{equation}
where $C_{(\cdot)} > 0$ are detection costs associated with the three different types of errors and $\pi_{(\cdot)}'$ are \emph{asserted} priors that may differ from the \emph{database priors} introduced in \eqref{eq:tandem-prior}. Both the detection costs and asserted priors are constants that one fixes prior to observing the evaluation data. Substituting the concurrent t-EER to \eqref{eq:tdcf-eq} gives
    \begin{equation}
        \text{t-DCF} = P_\mathscr{E}^\times\underbrace{\bigg(C_\text{miss}\pi_\text{tar} + C_\text{fa,non}\pi_\text{tar} + C_\text{fa,spf}\pi_\text{spf}\bigg)}_{\texttt{Constant}},\\
    \end{equation}
i.e.\ the t-DCF and the t-EER (at $\vec{\tau}_\times$) are linearly related. Furthermore, by denoting $C_\text{min}\mydef \min\big(C_\text{miss},C_\text{fa},C_\text{fa,spf}\big)$ and $C_\text{max}\mydef \max\big(C_\text{miss},C_\text{fa},C_\text{fa,spf}\big)$, the convexity of the expression in the parentheses (the asserted priors sum up to 1) can be used to sandwich the t-DCF value between lower and upper bounds:
\begin{equation}\label{eq:tDCF-bound}
    C_\text{min} \cdot P_\mathscr{E}^\times \leq \text{t-DCF} \leq C_\text{max}\cdot P_\mathscr{E}^\times. 
\end{equation}
These results indicate a close connection between the concurrent t-EER and the t-DCF.

\section{Experiments}\label{sec:experiments}

\begin{table}[]
\renewcommand{\arraystretch}{1.4}
\setlength\tabcolsep{3pt}
\caption{Number of target, non-target and spoof trials in the evaluation partitions of the SASV and ASVspoof 2021 logical access (LA) databases.  Due to the nature of the challenge in which participants prepared CM solutions alone, there are different protocols for the training of CMs and the ASV system for ASVspoof 2021 LA database.} 
\begin{tabular}{|ccc|ccccc|}
\hline
 \multicolumn{3}{|c|}{\multirow{2}{*}{\textbf{SASV}}}                                              & \multicolumn{5}{c|}{\textbf{ASVspoof 2021 LA}}                                                                                                                                \\ \cline{4-8} 
                                   \multicolumn{3}{|c|}{}                                                                            & \multicolumn{2}{c|}{\textbf{CM trials}}                                             & \multicolumn{3}{c|}{\textbf{ASV trials}}                                                                  \\ \hline
                                   \multicolumn{1}{|c|}{\textbf{target}} & \multicolumn{1}{c|}{\textbf{non-target}} & \textbf{spoof} & \multicolumn{1}{c|}{\textbf{target}} & \multicolumn{1}{c|}{\textbf{spoof}} & \multicolumn{1}{c|}{\textbf{target}} & \multicolumn{1}{c|}{\textbf{non-target}} & \textbf{spoof} \\ \hline
 \multicolumn{1}{|c|}{38697}           & \multicolumn{1}{c|}{33327}               & 63882          & \multicolumn{1}{c|}{14816}           & \multicolumn{1}{c|}{133360}         & \multicolumn{1}{c|}{13467}           & \multicolumn{1}{c|}{543114}              & 133362         \\ \hline
\end{tabular}
\label{tab: database details}
\end{table}

\begin{table*}[]
\renewcommand{\arraystretch}{1.4}
\setlength\tabcolsep{6pt}
\caption{
Equal error rates (EERs) for SASV and ASVspoof 2021 databases and baseline solutions.  
In each case, ASV EERs are conventional EERs estimated from a set of positive and negative class trials.  
Positive class trials are always target trials, whereas negative class trials involve a mix of non-target trials and spoofed trials where the proportions of each are dictated by the spoofing prevalence prior $\rho$.  
For each given $\rho$, the ASV system and hence also the EER is the same for all ASVspoof 2021 experiments. 
The CM EER is estimated from a mix of positive class (bona fide) trials and negative (spoofed) trials.  
The concurrent t-EER is estimated from the combination of ASV and CM sub-systems using the same trial sets.  
Tandem detection cost functions (t-DCF) results are provided for comparison.}
\begin{tabular}{|c||ccccc|c|ccccc|c|}
\hline

\multirow{2}{*}{\textbf{System}} & \multicolumn{5}{c|}{\textbf{ASV EER}}                                                                                                    & \multirow{2}{*}{\begin{tabular}[c]{@{}c@{}}\textbf{CM} \\ \textbf{EER}\end{tabular}} & \multicolumn{5}{c|}{\textbf{Concurrent t-EER}}                                                                                          & \multirow{2}{*}{\begin{tabular}[c]{@{}c@{}}\textbf{Min} \\ \textbf{t-DCF}\end{tabular}} \\ \cline{2-6} \cline{8-12}
& \multicolumn{1}{c|}{$\rho=0$} & \multicolumn{1}{c|}{$\rho=0.2$} & \multicolumn{1}{c|}{$\rho=0.5$} & \multicolumn{1}{c|}{$\rho=0.8$} & \multicolumn{1}{c|}{$\rho=1$} & & \multicolumn{1}{c|}{$\rho=0$} & \multicolumn{1}{c|}{$\rho=0.2$} & \multicolumn{1}{c|}{$\rho=0.5$} & \multicolumn{1}{c|}{$\rho=0.8$} & \multicolumn{1}{c|}{$\rho=1$} \\
\hline \hline
SASV baseline B1        & \multicolumn{1}{c|}{1.63}  & \multicolumn{1}{c|}{12.47}    & \multicolumn{1}{c|}{20.85}   & \multicolumn{1}{c|}{28.15}   & 30.74 & 0.67                                                               & \multicolumn{1}{c|}{2.28}  & \multicolumn{1}{c|}{2.28}    & \multicolumn{1}{c|}{2.28}   & \multicolumn{1}{c|}{2.28}    & 2.28  & 0.0660                                                                \\  \hline  \hline 
ASVspoof 2021 LA-B1      & \multicolumn{1}{c|}{7.61}  & \multicolumn{1}{c|}{14.56}   & \multicolumn{1}{c|}{25.16}   & \multicolumn{1}{c|}{34.01}   & 38.49 & 15.53                                                              & \multicolumn{1}{c|}{14.47} & \multicolumn{1}{c|}{14.47}   & \multicolumn{1}{c|}{14.47}  & \multicolumn{1}{c|}{14.47}   & 14.47 & 0.4930                                                                \\ \hline
ASVspoof 2021 LA-B2      & \multicolumn{1}{c|}{7.61}  & \multicolumn{1}{c|}{14.56}   & \multicolumn{1}{c|}{25.16}   & \multicolumn{1}{c|}{34.01}   & 38.49 & 19.97                                                              & \multicolumn{1}{c|}{18.42} & \multicolumn{1}{c|}{18.42}   & \multicolumn{1}{c|}{18.42}  & \multicolumn{1}{c|}{18.42}   & 18.42 & 0.5857                                                                \\ \hline
ASVspoof 2021 LA-B3      & \multicolumn{1}{c|}{7.61}  & \multicolumn{1}{c|}{14.56}   & \multicolumn{1}{c|}{25.16}   & \multicolumn{1}{c|}{34.01}   & 38.49 & 9.26                                                               & \multicolumn{1}{c|}{9.37}  & \multicolumn{1}{c|}{9.37}    & \multicolumn{1}{c|}{9.37}   & \multicolumn{1}{c|}{9.37}    & 9.37  & 0.3445                                                                \\ \hline
ASVspoof 2021 LA-B4      & \multicolumn{1}{c|}{7.61}  & \multicolumn{1}{c|}{14.56}   & \multicolumn{1}{c|}{25.16}   & \multicolumn{1}{c|}{34.01}   & 38.49 & 8.66                                                               & \multicolumn{1}{c|}{10.34} & \multicolumn{1}{c|}{10.34}   & \multicolumn{1}{c|}{10.34}  & \multicolumn{1}{c|}{10.34}   & 10.34 & 0.4027                                                                \\ \hline
\end{tabular}
\label{tab:results}
\end{table*}

To demonstrate use of the new t-EER using empirical scores, we conducted a set of experiments using SASV and ASVspoof databases and baseline systems.  Both SASV and ASVspoof 2021 logical access (LA) databases are sourced from the same ASVspoof 2019 database, while the baselines are different.

\subsection{Experimental setup}

SASV 2022 experiments were performed using the AASIST CM model described in~\cite{jung2022aasist} and the ECAPA-TDNN ASV model descibed in~\cite{desplanques2020ecapa}.  Together they are the B1 baseline described in~\cite{shim22_odyssey}. Experiments were performed using the publicly available SASV 2022 experimental protocol~\cite{jung2022sasv}.

We used all four ASVspoof 2021 baseline systems.  The ASV subsystem was fixed by the challenge organisers and is a standard \emph{x-vector}~\cite{snyder2018x} system with probabalistic linear discriminant analysis (PLDA) scoring~\cite{prince2007probabilistic}. The CM subsystems use: B1~-- constant-Q cepstral coefficients (CQCCs) and a Gaussian mixture model (GMM); B2~-- high resolution linear frequency cepstral coefficients (LFCCs) and a GMM; B3~-- LFCCs with a light convolutional neural network (LCNN); B4~-- a RawNet2 model.  
Further details of each are available in~\cite{yamagishi21_asvspoof}.  Experiments were performed using the ASVspoof 2021 experimental protocols~\cite{delgado2021_ASV_spoof,Liu2022ASVspoof2T}.

The number of target, non-target and spoofed trials for the evaluation partition of each database is shown in Table~\ref{tab: database details}.  For the ASVspoof 2021 LA challenge, participants prepared CM solutions alone, while the ASV system was designed by the challenge organisers.
There are hence different protocols for the evaluation of each.  
Since we are not concerned with optimisation in this work, all experiments were performed using the pretrained baseline systems with default settings and  the respective evaluation partitions of each database.

\subsection{Results}

Results are illustrated in Table~\ref{tab:results}.  
Columns~2-12 show EER results (including concurrent t-EER) for each database and baseline shown in column 1.  
Column~13 shows minimum tandem detection cost function (min~t-DCF) results for comparison.  
Columns~2-6 show ASV EER results for protocols comprising only: target and non-target trials 
($\rho=0$); target and spoofed trials ($\rho=1$); target vs.\ nontarget and spoofed trials for three intermediate spoofing prevalence priors ($0<\rho<1$). Column~7 shows the CM EER for a protocol comprising bona fide (target and nontarget) and spoofed trials. 

For a given database and $\rho$, the ASV EERs are the same, no matter what the baseline.
They vary between 1.6\% and 30.7\% for the SASV database and between 7.6\% and 38.5\% for the ASVspoof database.  Unsurprisingly, as the prevalence of spoofed trials increases (increasing $\rho$), so too does the EER. The dependence of the EER upon the spoofing prevalence prior is undesirable.  
Experiments involving different values of $\rho$ are, in effect, akin to experiments using different databases and protocols with different spoofing prevalence priors. The differences imply that performance comparisons across databases cannot be made using the EER.  

Columns~8-12 show corresponding concurrent t-EER results for the same range of $\rho$.  They are close to being identical for each row, indicating that they are dependent on the system, but not the spoofing prevalence prior (nor any other prior).  The independence of the concurrent t-EER to priors implies that it can be used as a more reliable metric for making performance comparisons across different databases.  

Plots similar to those in Fig.~\ref{fig:tEER_example_2d}, but for empirically-derived scores for the SASV B1 baseline, are shown in Fig.~\ref{fig:tEER_example_2d_real_scores} for the same values of $\rho$ shown in Table~\ref{tab:results}. The EER for the $\rho=1$ condition (no nontargets) of 30.7\% corresponds to the t-EER value for the highest ASV threshold $\tau_{\text{asv}}$ (right-most points in Fig.~\ref{fig:tEER_example_2d_real_scores}(b)).  The EER for the ($\rho=0$) condition (no spoofs) of 1.6\% furthermore corresponds to the lowest point on the t-EER path for $\rho=0$.  The CM EER of 0.7\% is shown by the horizontal asymptote of the t-EER path for $\rho=1$.  Finally, the concurrent t-EER of 2.3\% is shown by the magenta marker.

While not the goal in this paper, the \emph{ranking} of concurrent t-EER results reflects the same trend as min t-DCF results.  In the case of results for the ASVspoof 2021 database, the ranking derived from CM EER results alone (ASV results are in any case invariable across different CM solutions for a given $\rho$) is different.  
It might seem at first that B4 is the best among the four baselines (the CM EER is lowest).  Tandem results, however, show that B3 might be a better option (the t-EER is lowest).  This observation adds further weight to the argument that the selection of CM solutions should not be made from CM results alone, but instead be made from the results of \emph{tandem assessment}.  The concurrent t-EER offers one suitable approach.

\section{Discussion and Conclusions}\label{sec:conclusions}

\subsection{Summary of the Proposed t-EER Metric}

Experimental results reported in this paper corroborate previous findings that the evaluation of spoofing countermeasure solutions should be performed in tandem with biometric recognisers; their evaluation in isolation might result in the selection of a sub-optimal solution. By extending the standard notion of detection theory and the EER, we introduced an approach to tandem evaluation in the form of the tandem equal error rate (t-EER). Its application results in a path of operating points --- the t-EER path --- along which the tandem false alarm and miss rates are equal. Nonetheless, each operating point along each path corresponds to a different t-EER value and the path itself is dependent upon the prevalence of spoofing attacks in the database. We tackled this problem by introducing \emph{concurrent} t-EER operating point at which all the paths intersect, thereby giving a unique error rate that is prior-independent. Just as the conventional EER corresponds to an operating point at which the errors associated with the two classes are equal ($P_\text{miss}=P_\text{fa}$), the concurrent t-EER corresponds to an operating point at which the three \emph{tandem} error rates are equal: $P_\text{miss}^\text{tdm}=P_\text{fa}^\text{tdm}=P_\text{fa,spoof}^\text{tdm}$. 

Besides the concurrent t-EER point, an intuitive error-rate based performance measure of the tandem system, our work also sheds light upon the various per-system EERs used in the separate evaluation of biometric comparator and PAD subsystems. The t-EER path also reveals the: 
    \begin{itemize}
        \item EER of the biometric comparator in the absence of spoofing attacks (an accept-all PAD, and spoofing attack prior $\pi_\text{spoof}=0$);
        \item EER of the biometric comparator in the absence of non-target trials (an accept-all PAD, spoofing attack prior $\pi_\text{spoof}=1$);
        \item EER of the biometric comparator with a mix of nontargets and spoofing attacks (an accept-all PAD, and spoofing attack $(0 < \pi_\text{spoof} < 1$);
        \item EER of the PAD subsystem (accept-all comparator).
    \end{itemize}
The t-EER path hence presents a \emph{unified} view of biometric comparator and PAD subsystem performance under a variety of protocols and configurations. The authors advocate for the adoption of the t-EER framework in place of traditional two-class EERs when assessment involves three classes -- target, non-target and spoofed trials. We hence discourage the continued use of two-class EER metrics, such as the SASV-EER adopted for the SASV challenge~\cite{SASV2022-eval-plan}; it is dependent on the database prevalence prior and potentially leaves comparisons of performance made across different datasets essentially meaningless.

\subsection{Towards a reference approach to joint evaluation}

Though an awareness of vulnerabilities to PAs was founded some two decades ago \cite{Ratha2001-enhancing} and though a framework for PAD research and performance assessment has been standardized~\cite{ISOpresentationAtack}, research into the \emph{joint} operation and evaluation of biometric comparator and PAD subsystems remains in its relative infancy.  The usual approach, score fusion, transforms the task of joint \emph{evaluation} into one of \emph{back-end classifier design} with nearly limitless design possibilities and inherent training data dependencies.

Being convinced that simplicity is the key towards defining a reference approach, the authors have advocated a natural two-threshold, three-class extension of single-system ROC analysis to support the tandem evaluation of biometric comparators and PAD subsystems. The main assumption (the same as that in~\cite{Kinnunen2020-tandem-fundamentals}) is the statistical independence of the scores produced by each subsystem. It is justified by noting that PAD and biometric comparators address  different classification tasks. We hope the work will inspire further discussion on tandem evaluation through a simple error-rate based approach. By avoiding the need to specify application-specific parameters, we embrace a spirit more in keeping with the ISO standardised approach to the separate evaluation of PAD. The new t-EER metric can be used for the comparison of different solutions and may also be applied when making comparisons between results derived from different databases. Importantly, the metric is agnostic to the biometric mode and requires only scores produced by biometric comparators and PAD subsystems. Hence, the proposed metric has potential to be adapted as a common reference approach to joint evaluation. To this end, the authors provide an open-source Python implementation\textsuperscript{\ref{fn:url}}.

A potential interesting future extension concerns tandem evaluation of \emph{more than two systems}. Biometric systems are known to be vulnerable to many different kinds of attack vectors (e.g. replay and text-to-speech for speech) that typically benefit from tailored countermeasures combined with the biometric comparator. An even more general situation involves multiple biometric modes (e.g. face and voice), each with potentially multiple biometric comparators and/or PAD subsystems. We are again faced with the key question how the different systems should be combined. Assuming that all the relevant subsystems can be collapsed into a single score (e.g. score fusion of multiple PADs), the t-EER approach remains applicable. More principled tandem assessment of multiple PADs and biometric modes is left as a future work.

\subsection{Fundamental Limitations of (t-)EER}

The present work should not be viewed as the end but, rather, a first rigorous step towards the unified joint evaluation of biometric comparators and PAD subsystems. The authors are well aware that our starting point --- the classical EER --- has a number of known limitations, namely:
    \begin{itemize}
        \item the EER threshold is selected by an oracle using the ground-truth class labels of the particular evaluation database, hence the problems of threshold selection or the training of a calibration model is not accounted for; 
        \item an operating point specified by balanced errors (misses and false alarms) will not suit all applications where preference might naturally be given to lowering the miss rate or the false alarm rate.
    \end{itemize}
The first shortcoming is exactly why the community \cite{bengio04b-EPC,Chingovska2014-biometrics-eval-under-spoofing-attacks,BRUMMER2006-application-independent,Ferrer2022-robust-calibration,Hand2021-notes-on-hmeasure} encourages assessment approaches that take threshold selection into consideration from the very beginning. The second accounts for why metrics such as the DCF~\cite{DODDINGTON2000-NIST} and t-DCF~\cite{Kinnunen2020-tandem-fundamentals} require the setting of detection costs and asserted priors. Being an extension of the classical EER, the proposed t-EER inherits both of the above shortcomings. So, why did we bother to write this paper?

The main reason is that, \textbf{despite the acknowledged drawbacks, the EER remains a popular metric for evaluation and reporting}. As a matter of fact, there is nothing particularly wrong in using the EER as a \emph{discrimination} (class separability) criterion, similar to the popular AUC metric --- as long as the above fundamental limitations are understood and acknowledged. The lesser-known interpretation of the EER as an upper bound to the total error rate~\cite{BrummerFS21} indeed makes the EER well suited to measurements of class separability. Like the AUC, the EER is useful for quick and easily interpreted analyses and for initial investigations, such as \emph{rejecting} features or models that lack the potential for accurate classification. That the EER is easily visualized on a ROC or DET plots, that it is independent of the class proportions, and that is can be estimated without the need for additional data beyond the evaluation data are all likely reasons for why the EER remains popular.

While calibration remains an important (yet different) problem, the focus in this paper is upon discrimination and a proposal towards a common approach to the joint evaluation of biometric comparators and PAD subsystems built upon foundations that should be familiar to anyone working or having an interest in biometric recognition. In fact, given the lack of any candidate approaches in the current ISO/IEC PAD standard~\cite{ISOpresentationAtack}, \emph{any} reasonable metric provides a suitable starting point for the future definition of a common approach to the joint evaluation of biometric comparators operating in tandem with PAD subsystems.  Ultimately, we hope the new t-EER serves also as an eventual starting point for the exploration of approaches to the tandem evaluation of calibration in addition to discrimination.

\section*{Acknowledgment}

The work has been partially supported by the Academy of Finland (Decision No.\ 349605, project ``SPEECHFAKES''), the Agency of Science, Technology and Research (A$^\star$STAR), Singapore, through its CRF Core Project Scheme (Project No.\ CR-2021-005) and by the Agence Nationale de la Recherche (ANR), France, through VoicePersonae (Project No.\ ANR-18-JSTS-0001).

\ifCLASSOPTIONcaptionsoff
  \newpage
\fi



%


\bibliographystyle{IEEEtran}
\bibliography{bibliography}

%

\begin{IEEEbiography}[{\includegraphics[width=1in,height=1.25in,clip,keepaspectratio]{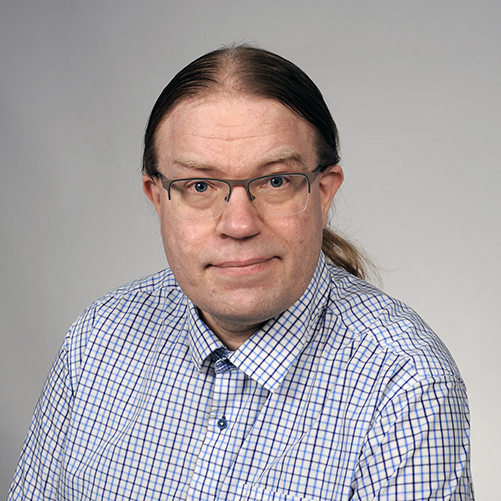}}]{Tomi H. Kinnunen} is full professor of speech technology at the University of Eastern Finland (UEF). He received his Ph.D. degree (computer science) from the University of Joensuu in 2005. From 2005 to 2007, he was with the Institute for Infocomm Research (I2R), Singapore. Since 2007, he has been with UEF. His research has been funded by Academy of Finland and he also partnered in H2020-funded OCTAVE project. He chaired the Odyssey workshop in 2014. From 2015 to 2018, he served as an Associate Editor for IEEE/ACM-T-ASLP and from 2016 to 2018 as a Subject Editor in Speech Communication. In 2015 and 2016, he visited the National Institute of Informatics, Japan, for 6 months under a mobility grant from the Academy of Finland. He is one of the co-founders of the ASVspoof challenge, a nonprofit initiative that seeks to evaluate and enhance the security of voice biometric solutions under spoofing attacks.
\end{IEEEbiography}

\begin{IEEEbiography}
[{\includegraphics[width=1.2in,height=1.0in,clip,keepaspectratio]{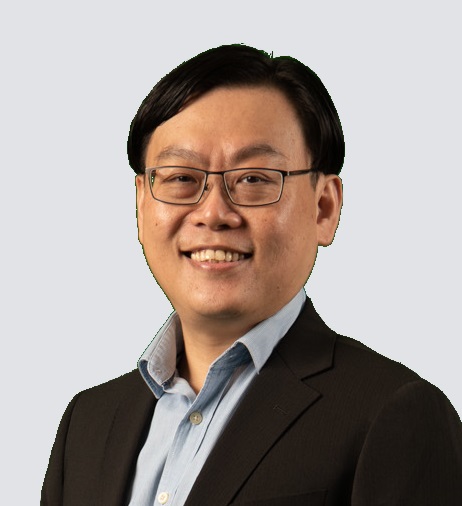}}]
{Kong Aik Lee} (M'05-SM'16) is currently an Associate Professor with the Singapore Institute of Technology, Singapore. He holds a concurrent appointment as a Principal Scientist and a Group Leader with the Agency for Science, Technology and Research (A $^\star$ STAR), Singapore. From 2018 to 2020, he was a Senior Principal Researcher with the Data Science Research Laboratories, NEC Corporation, Tokyo, Japan. He received his Ph.D. from Nanyang Technological University, Singapore, in 2006. After which, he joined the Institute for Infocomm Research, Singapore, as a Research Scientist and then a Strategic Planning Manager (concurrent appointment). His research interests include the automatic and para-linguistic analysis of speaker characteristics, ranging from speaker recognition, language and accent recognition, diarization, voice biometrics, spoofing, and countermeasure. He was the recipient of the Singapore IES Prestigious Engineering Achievement Award 2013 for his contribution to voice biometrics technology, and Outstanding Service Award by IEEE ICME 2020. Since 2016, he has been an Editorial Board Member of Elsevier Computer Speech and Language. From 2017 to 2021, he was an Associate Editor for IEEE/ACM Transactions on Audio, Speech, and Language Processing. He is an elected Member of the IEEE Speech and Language Processing Technical Committee and was the General Chair of the Speaker Odyssey 2020 Workshop.
\end{IEEEbiography}

\begin{IEEEbiography}[{\includegraphics[width=1in,height=1.25in,clip,keepaspectratio]{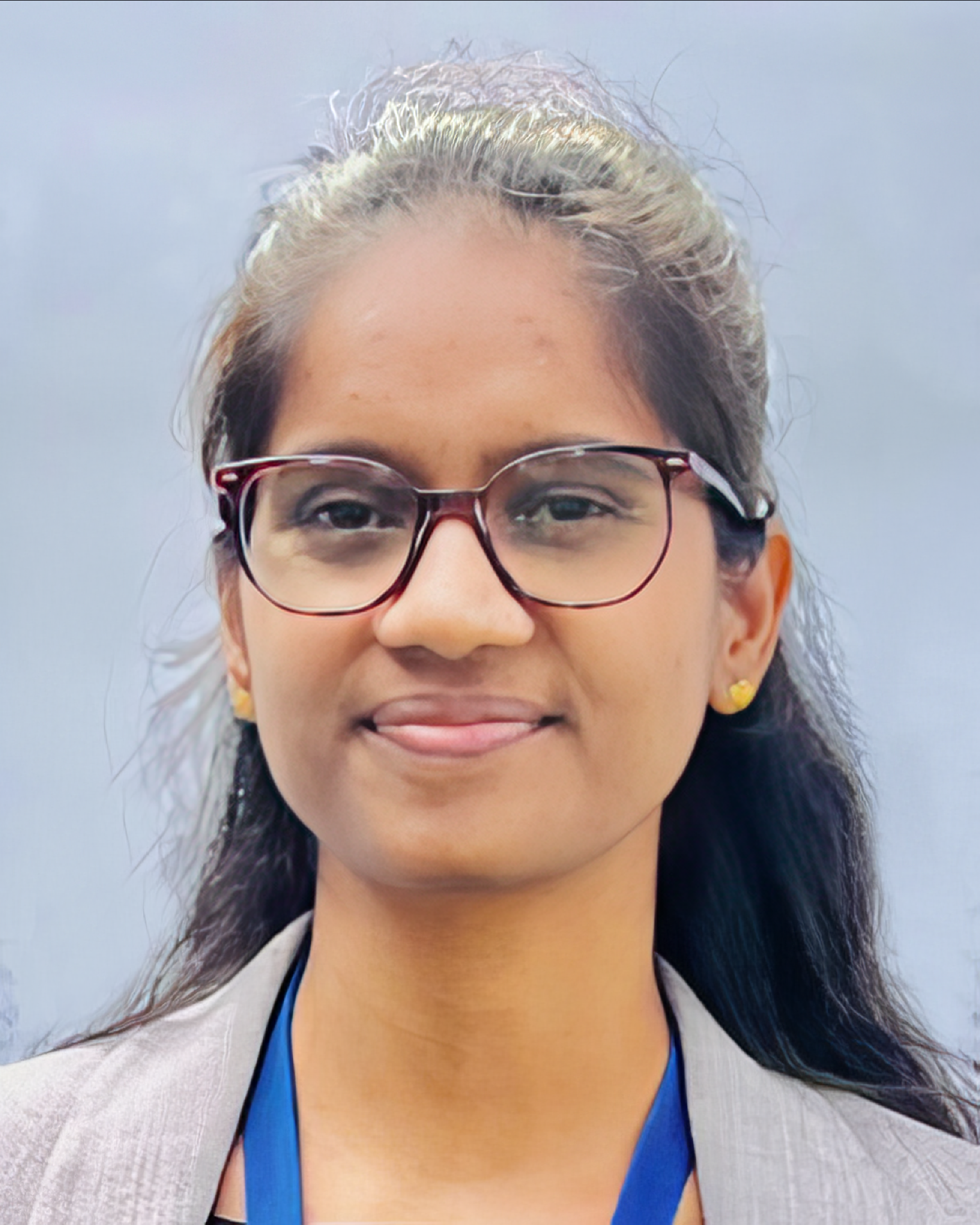}}]{Hemlata Tak} received her Ph.D. degree from Sorbonne University, France. She received her Master’s degree in 2018 from DA-IICT, Gandhinagar, India. She co-organized the inaugural edition of the Spoof-Aware Speaker Verification (SASV) Challenge 2022. She is also a co-organiser of the ASVspoof 5 Challenge. Her research interests include voice biometrics, audio deepfake detection and anti-spoofing.
\end{IEEEbiography}

\begin{IEEEbiography}[{\includegraphics[width=1in,height=1.25in,clip,keepaspectratio]{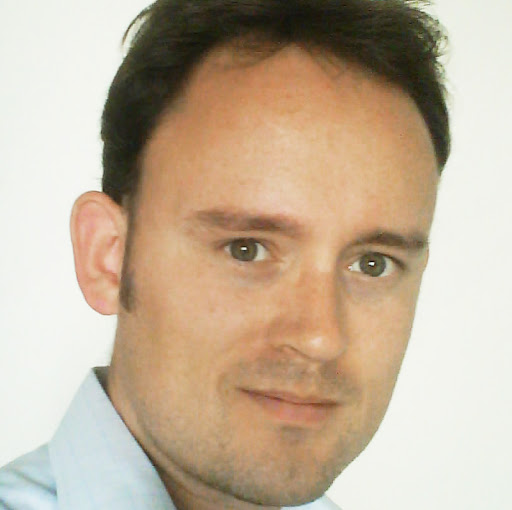}}]{Nicholas Evans} (Member, IEEE) is a Professor at EURECOM, France, where he heads research in Audio Security and Privacy. He received his Ph.D.\ from the University of Wales Swansea, UK in 2004. He is a co-founder of the community led, ASVspoof, SASV, and VoicePrivacy challenge series. He participated in the EU FP7 Tabula Rasa and EU H2020 OCTAVE projects, both involving antispoofing. Today, his team is leading the EU H2020 TReSPAsS-ETN project, a training initiative in security and privacy for multiple biometric characteristics. He co-edited the second and third editions of the Handbook of Biometric Anti-Spoofing, served previously on the IEEE Speech and Language Technical Committee and serves currently as an associate editor for the IEEE Trans.\ on Biometrics, Behavior, and Identity Science.
\end{IEEEbiography}

\begin{IEEEbiography}[{\includegraphics[width=1in,height=1.25in,clip,keepaspectratio]{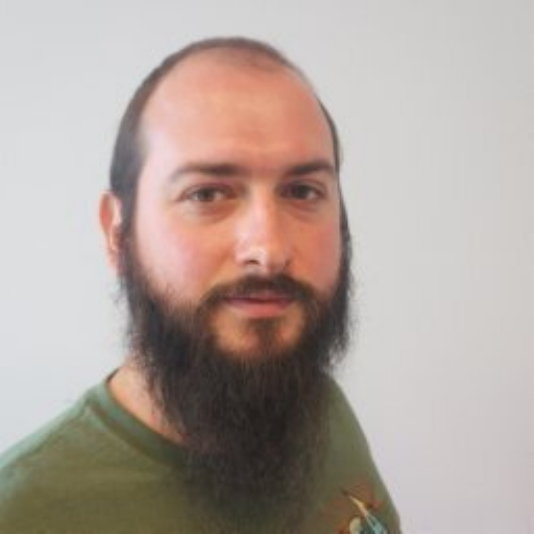}}]{Andreas Nautsch} was research engineer at Universit\'{e} d'Avignon and senior research fellow at EURECOM. He received his doctorate from TU Darmstadt (2019), served as project editor for ISO/IEC 19794-13, co-organised editions of the VoicePrivacy and ASVspoof challenges, co-initiated the ISCA SIG on Security \& Privacy in Speech Communication (SPSC), was an Associate Editor for the EURASIP Journal on Audio, Speech, and Music Processing, co-lead the 2021 Lorentz workshop on Speech as Personal Identifiable Information, and co-maintained the OpenSource project SpeechBrain. By 2020, he lead multidisciplinary publication teams composed of speech \& language technologists, legal scholars, cryptographers, and biometric experts; Andreas co-responded on behalf of ISCA SIG-SPSC to the public consultation of the 2021 EDPB guidelines on virtual voice assistants. In 2023, he joined goSmart to lead the solution development (architecture, design, and implementation) for private 5G based campus networks.
\end{IEEEbiography}

\clearpage
\appendices
\numberwithin{equation}{section}
\renewcommand{\theequation}{\thesection.\arabic{equation}}
\pagestyle{empty}

\counterwithin{table}{section}
\renewcommand{\thetable}{\thesection.\arabic{table}}

\section{Proofs}
\subsection*{Proof of Lemma 5.2}
Note that the tandem miss rate 
\begin{equation}
P_\text{miss}^\text{tdm}(\tau_\text{cm})=\underbrace{P_\text{miss}^\text{asv}(\tau^\text{asv}_*)}_{\texttt{Constant}}+\underbrace{\big(1-P_\text{miss}^\text{asv}(\tau^\text{asv}_*)\big)}_{\texttt{Constant}}P_\text{miss}^\text{cm}(\tau_\text{cm})
\end{equation}
is a monotonically non-decreasing function of $\tau_\text{cm}$ for which $P_\text{miss}^\text{tdm}(-\infty)=P_\text{miss}^\text{asv}(\tau^\text{asv}_*)$  and $P_\text{miss}^\text{tdm}(+\infty)=1$. Likewise, the tandem false alarm rate 
    \begin{equation}
        \begin{aligned}
            P_\text{fa}^\text{tdm}(\tau_\text{cm}) & = (1 - \rho)\left[(1 - P_\text{miss}^\text{cm}(\tau_\text{cm}) \right] \cdot \underbrace{P_\text{fa,non}^\text{asv}(\tau^\text{asv}_*)}_{\texttt{Constant}}\nonumber\\
            & + \rho P_\text{fa}^\text{cm}(\tau_\text{cm})\underbrace{P_\text{fa,spoof}^\text{asv}(\tau^\text{asv}_*)}_{\texttt{Constant}}
        \end{aligned}
    \end{equation}
is easily seen to be monotonically non-increasing function of $\tau_\text{cm}$, with  $P_\text{fa}^\text{tdm}(-\infty)=(1-\rho)P_\text{fa,non}^\text{asv}(\tau^\text{asv}_*) + \rho P_\text{fa,spoof}^\text{asv}(\tau^\text{asv}_*)$ and $P_\text{fa}^\text{tdm}(+\infty)=0$. The existence of tandem EER operating point is equivalent of stating that the graphs of $P_\text{miss}^\text{tdm}(\tau_\text{cm})$ and $P_\text{fa}^\text{tdm}(\tau_\text{cm})$ intersect. This is possible if (and only if) the condition \eqref{eq:asv-feasible-region} 
holds. $\blacksquare$

\subsection*{Proof of Theorem 6.1}
By the definition of t-EER path,
    \begin{equation}
        \begin{aligned}
        \Omega(\rho_1) & = \{ \vec{\tau} \in \mathbb{R}^2: P_\text{miss}^\text{tdm}(\vec{\tau}) = P_\text{fa,$\rho_1$}^\text{tdm}(\vec{\tau})\}\nonumber\\
        \Omega(\rho_2) & = \{ \vec{\tau} \in \mathbb{R}^2: P_\text{miss}^\text{tdm}(\vec{\tau}) = P_\text{fa,$\rho_2$}^\text{tdm}(\vec{\tau})\}\nonumber.
        \end{aligned}
    \end{equation}
From the definition of the intersection, $\vec{\tau}^\times \in  \Omega(\rho_1) \cap  \Omega(\rho_2)$, i.e. $\vec{\tau}^\times \in  \Omega(\rho_1)$ \textbf{and} $\vec{\tau}^\times \in  \Omega(\rho_2)$, i.e. the following must be fulfilled simultaneously:
    \begin{equation}
        \begin{aligned}
            P_\text{miss}^\text{tdm}(\vec{\tau}^\times) & = P_\text{fa,$\rho_1$}^\text{tdm}(\vec{\tau}^\times)\nonumber\\
            P_\text{miss}^\text{tdm}(\vec{\tau}^\times) & = P_\text{fa,$\rho_2$}^\text{tdm}(\vec{\tau}^\times),
        \end{aligned}
    \end{equation}
or 
    \begin{equation}\label{eq:eliminated-tdm-miss-rate}
        P_\text{fa,$\rho_1$}^\text{tdm}(\vec{\tau}^\times) = P_\text{fa,$\rho_2$}^\text{tdm}(\vec{\tau}^\times).       
    \end{equation}
By expanding \eqref{eq:eliminated-tdm-miss-rate} using the definition of the total false alarm rate \eqref{eq:total-tandem-fa-rate}, 
we have that
\begin{equation}
    (\rho_1 - \rho_2)\Bigg[P_\text{fa,non}^\text{tdm}(\vec{\tau}^\times) - P_\text{fa,spf}^\text{tdm}(\vec{\tau}^\times) \Bigg]=0,
    \label{eq:crossover}
\end{equation}
which is trivially fulfilled when $\rho_1=\rho_2$ (the case of identical t-EER paths). Since we however assumed $\rho_1 \neq \rho_2$, the zero-product property implies that the expression enclosed in the bracketed parentheses of \eqref{eq:crossover} must equal 0. Upon substituting the tandem error rates with their component error rates evaluated on the ASV and CM scores (defined in Section \ref{subsec:tandem-miss-and-FA-AND-rule}), 
we arrive at  
    \begin{equation}
       P_\text{fa,non}^\text{asv}(\tau_\text{asv}^\times)(1 - P_\text{miss}^\text{cm}(\tau_\text{cm}^\times)) - P_\text{fa}^\text{cm}(\tau_\text{cm}^\times)P_\text{fa,spf}^\text{asv}(\tau_\text{asv}^\times)=0,\nonumber
    \end{equation}
Equation \eqref{eq:xpoint-equation} follows.

The \emph{value} of the t-EER can then be read either from the tandem miss or the tandem false alarm rate at $\vec{\tau}_\times$ (the two are equal, by construction). To obtain the simple expression in (24) 
we consider the tandem false alarm rate \eqref{eq:total-tandem-fa-rate}. 
In particular, the cross-multiplying of \eqref{eq:xpoint-equation} 
reveals that $(1 - P_\text{miss}^\text{cm}(\tau_\text{cm}^\times))P_\text{fa,non}^\text{asv}(\tau_\text{asv}^\times)$ can be replaced by $P_\text{fa,spf}^\text{asv}(\tau_\text{asv}^\times)P_\text{fa}^\text{cm}(\tau_\text{cm}^\times)$. Equation 
\eqref{eq:t-EER-at-xpoint} follows. $\blacksquare$

\section{Additional analyses and experiments}

\begin{table}[htp]
\caption{Equal error rates (EERs) using different ASV and CM system combinations.  
In each case, ASV EERs are conventional EERs estimated from a set of positive (target) and negative (non-target) class trials.  
The CM EER is estimated from a mix of positive class (target) trials and negative (spoofed) trials. The concurrent t-EER is estimated from the combination of ASV and CM sub-systems using the same trial sets.}
\setlength\tabcolsep{9pt}
\begin{tabular}{|cc|c|c|c|}
\hline
\multicolumn{2}{|c|}{System}              & \multirow{2}{*}{\begin{tabular}[c]{@{}c@{}}ASV EER\\ (\%)\end{tabular}} & \multirow{2}{*}{\begin{tabular}[c]{@{}c@{}}CM EER\\ (\%)\end{tabular}} & \multirow{2}{*}{\begin{tabular}[c]{@{}c@{}}t-EER\\ (\%)\end{tabular}} \\ \cline{1-2}
\multicolumn{1}{|c|}{ASV}       & CM      &                                                                        &                                                                        &                                                                       \\ \hline
\multicolumn{1}{|c|}{ResNetSE34}  & RawNet2 & 1.27                                                                   & 3.41                                                                    & 3.69                                                                  \\ \hline
\multicolumn{1}{|c|}{ResNetSE34}  & AASIST  & 1.27                                                                   & 0.65                                                                   & 1.78                                                                  \\ \hline
\multicolumn{1}{|c|}{ResNetSE34}  & SSL     & 1.27                                                                   & 1.08                                                                   & 1.53                                                                  \\ \hline
\multicolumn{1}{|c|}{ECAPA}     & RawNet2 & 1.63                                                                   & 3.41                                                                    & 4.12                                                                  \\ \hline
\multicolumn{1}{|c|}{ECAPA}     & AASIST  & 1.63                                                                   & 0.65                                                                   & 2.28                                                                  \\ \hline
\multicolumn{1}{|c|}{ECAPA}     & SSL     & 1.63                                                                   & 1.08                                                                  & 1.88                                                                  \\ \hline
\multicolumn{1}{|c|}{SKA-TDNN}  & RawNet2 & 0.39                                                                   & 3.41                                                                    & 3.58                                                                  \\ \hline
\multicolumn{1}{|c|}{SKA-TDNN}  & AASIST  & 0.39                                                                   & 0.65                                                                   & 1.18                                                                  \\ \hline
\multicolumn{1}{|c|}{SKA-TDNN}  & SSL     & 0.39                                                                   & 1.08                                                                   & 0.95                                                                  \\ \hline
\multicolumn{1}{|c|}{Conformer} & RawNet2 & 0.41                                                                   & 3.41                                                                    & 3.67                                                                  \\ \hline
\multicolumn{1}{|c|}{Conformer} & AASIST  & 0.41                                                                   & 0.65                                                                   & 1.22                                                                  \\ \hline
\multicolumn{1}{|c|}{Conformer} & SSL     & 0.41                                                                   & 1.08                                                                   & 0.99                                                                  \\ \hline
\multicolumn{1}{|c|}{X-vector}  & RawNet2 & 2.46                                                                   & 3.41                                                                    & 4.45                                                                  \\ \hline
\multicolumn{1}{|c|}{X-vector}  & AASIST  & 2.46                                                                   & 0.65                                                                   & 2.78                                                                  \\ \hline
\multicolumn{1}{|c|}{X-vector}  & SSL     & 2.46                                                                   & 1.08                                                                   & 2.64                                                                  \\ \hline
\end{tabular}
\label{tab:additional_results}
\end{table}

\begin{table*}[htbp]
\setlength\tabcolsep{0.7pt}
    \centering
    \caption{Class-conditional correlations coefficients for all three different classes. ASV sub-systems are ASV1: ResNet34, ASV2: ECAPA, ASV3: SKA-TDNN, ASV4: Conformer  and ASV5: X-vector. CM sub-systems are: CM1: RawNet2, CM2: AASIST and CM3: SSL.} 
    \renewcommand{\arraystretch}{1.5}
    \begin{subtable}[t]{0.31\textwidth}
        \centering
        \begin{tabular}{|l|c|c|c|c|c|}
\hline\hline
            &ASV1 & ASV2 & ASV3&ASV4&ASV5 \\
            \hline
            CM1&0.317&0.287&0.090&0.091&0.057\\\hline
            CM2&0.237&0.198&0.088&0.102&0.063\\\hline
            CM3&0.205&0.173&0.065&0.081&0.057\\
            
            \hline
        \end{tabular}
        \caption{Target class}
        \label{subtable1}
    \end{subtable}
    \hfill
    \begin{subtable}[t]{0.31\textwidth}
        \centering
        \begin{tabular}{|l|c|c|c|c|c|}
\hline\hline
            &ASV1 & ASV2 & ASV3&ASV4&ASV5 \\
            \hline
            CM1&0.108&0.110&0.041&0.068&0.070\\\hline
           CM2&0.109&0.111&0.076&0.085&0.048\\\hline
           CM3&0.044&0.036&0.011&0.020&-0.033\\
            
            \hline
        \end{tabular}
        \caption{Non-target class}
        \label{subtable2}
    \end{subtable}
    \hfill
    \begin{subtable}[t]{0.31\textwidth}
        \centering
        \begin{tabular}{|l|c|c|c|c|c|}
\hline\hline
            &ASV1 & ASV2 & ASV3&ASV4&ASV5 \\
            \hline
            CM1&-0.360&-0.395&-0.363&-0.387&-0.402\\\hline
            CM2&0.367&0.308&0.345&0.321&0.272\\\hline
            CM3&-0.162&-0.190&-0.146&-0.175&-0.259\\
            
            \hline
        \end{tabular}
        \caption{Spoof class}
        \label{subtable3}
    \end{subtable}
    \label{tab:correlation-values-without-softmax}
\end{table*}

\begin{table}[!t]
    \centering
    \caption{Attack-wise correlations  ASV (ECAPA) and CM (AASIST) scores in the SASV task.} \begin{tabular}{c|c}
        Attack id & Correlation\\
        \hline\hline
        A07 & 0.343\\
        A08 & 0.246\\
        A09 & -0.120\\
        A10 & 0.342\\
        A11 & 0.221\\
        A12 & 0.470\\
        A13 & 0.225\\
        A14 & 0.062\\
        A15 & 0.266\\
        A16 & 0.141\\
        A17 & 0.065\\
        A18 & 0.090\\
        A19 & 0.002\\
        \hline 
    \end{tabular}
    \label{tab:attack-wise-correlation}
\end{table}

\subsection{Additional state-of-the-art recognizers}

We conducted additional experiments using combinations of different state-of-the-art ASV and CM sub-systems to compute the traditional ASV EER (target \textit{vs} non-target trials), CM EER (target \textit{vs} spoof trials), and concurrent t-EER derived from the combined ASV and CM sub-systems using the same trial sets. These experiments were performed using the publicly available SASV 2022 experimental protocol~\cite{jung2022sasv}. In this study, we explored several publicly available ASV systems, including ResNetSE34~\cite{kwon2021ins}, ECAPA~\cite{desplanques2020ecapa}, a selective kernel attention-based time delay neural network (SKA-TDNN)\cite{mun2023frequency}, a multi-scale feature aggregation conformer (MFA-Conformer)\cite{zhang2022mfa}, and X-vector~\cite{snyder2018x} with probabilistic linear discriminant analysis (PLDA) scoring~\cite{prince2007probabilistic}. Similarly, we explored various CM models, such as AASIST~\cite{jung2022aasist}, a self-supervised front-end based CM system~\cite{tak2022automatic,wang2021investigating}, and RawNet2~\cite{tak2021end}.

Results are illustrated in Table~\ref{tab:additional_results}. We selected ASV and CM sub-systems with varied degree of complexity and accuracy with following combinations i) lower-EER detector \textit{vs} higher accuracy recogniser; ii) lower-EER detector \textit{vs} lower accuracy recogniser; iii) higher-EER detector \textit{vs} lower accuracy recogniser; iv) higher-EER detector \textit{vs} higher accuracy recogniser. Regardless of the choice of the CM sub-systems, the ASV EERs remain constant across rows 2 to 4, as well as rows 5 to 7, 8 to 10, 11 to 13, and the last 14 to 16 rows. Last column shows the corresponding concurrent t-EER results for the different combination of ASV and CM sub-systems. The t-eer values vary between 0.95\% and 4.45\%. The best two sub-system in the lowest t-EER combination are SKA-TDNN+SSL with a t-EER value of 0.95\%, whereas using the individually-best CM (AASIST) and ASV (SKA-TDNN) sub-system with lowest EER combination results in a t-EER value of 1.18\%. The fact that the concurrent t-EER is independent of the best ASV and CM sub-systems implies that it can be utilised as a comprehensive evaluation metric.

\subsection{Cross-correlation of ASV and CM}
\label{sec:correlation}
Table~\ref{tab:correlation-values-without-softmax} shows the class-conditional correlations between ASV and CM systems. As can be seen, the degree of correlation depends on the pair of systems under consideration. In this data, we observe overall lower correlations for the non-target class compared to the other two classes. Additionally, we display \emph{per-attack} correlations for an arbitrary pair of systems in Table \ref{tab:attack-wise-correlation}. As can be seen, the correlation depends on the spoofing attack; some attacks (e.g., attack A12) exhibit a high degree correlation, whereas others (e.g., attacks A17---A19) little to no correlation.

The class-conditional independence assumption made in Subsection \ref{subsec:tandem-miss-and-FA-AND-rule} is common to both the t-DCF metric \cite{Kinnunen2020-tandem-fundamentals} and the t-EER put forward in this paper. It is similar to other independence assumptions made in modeling and performance evaluation (e.g., treating test trials originating from the same biometric capture subject as if they were `independent', even though statistical dependency is introduced through the subject identity). As in any modeling task, the validity of assumptions should be scrutinized whenever in doubt. \emph{Despite} observing non-neglibgle ASV and CM correlations for certain pairs of ASV and CM systems (Table \ref{tab:correlation-values-without-softmax}) or specific attacks (Table \ref{tab:attack-wise-correlation}), our experiments demonstrate that the results obtained using the t-EER as an evaluation metric are consistent with what might be expected from per-system EERs.











\end{document}